\documentclass[aps,pra,reprint,groupedaddress]{revtex4-1}
\usepackage{graphicx}
\usepackage{amsmath}
\usepackage{amssymb}
\usepackage{xcolor}
\begin{document}


\title{Electromagnetic wave propagation through a slab of a dispersive medium}

\author{Mohamed Ismail}
\affiliation{Aix Marseille Univ, CNRS, Centrale Marseille, 
Institut Fresnel, Marseille, France}
\author{Boris Gralak}
\affiliation{Aix Marseille Univ, CNRS, Centrale Marseille, 
Institut Fresnel, Marseille, France}

\def\ep{\varepsilon}
\def\eps{\varepsilon_s}
\def\om{\omega}
\def\oms{\om_s}
\def\omr{\om_0}
\def\omp{\om_p}
\def\omq{\om_q}
\def\Om{\Omega}
\def\hom{\hat\omega}
\def\homs{\hat\om_s}
\def\homr{\hat\om_0}
\def\homp{\hat\om_p}
\def\homq{\hat\om_q}
\def\hOm{\hat\Omega}
\def\r{r}
\def\xs{x_0}
\def\phase{\phi}
\def\R{R}
\def\T{T}

\def\b{\color{blue}}
\def\br{\color{red}}


\begin{abstract}
A method is proposed for the analysis of the propagation of electromagnetic waves 
through a homogeneous slab of a medium with Drude-Lorentz dispersion behavior, and excited 
by a causal sinusoidal source. An expression of the time dependent field, free from branch-cuts in the 
plane of complex frequencies, is established. This method provides the complete temporal response 
in both the steady-state and transient regimes in terms of discrete poles contributions. 
The Sommerfeld and Brillouin precursors are retrieved and the corresponding set of poles are identified. 
In addition, the contribution in the transient field of the resonance frequency in the Drude-Lorentz 
model is exhybited, and the effect of reflections resulting from the refractive index mismatch 
at the interfaces of the slab are analyzed. 
\end{abstract}


\maketitle

\section{Introduction\label{intro}}

Propagation of electromagnetic waves in dispersive homogeneous media has been originally 
investigated by Brillouin and Sommerfeld \cite{Bri1960}. 
More recently, the introduction of negative index materials 
\cite{Ves68}, metamaterials \cite{Pendry00,Notomi00,Gralak00} and 
transformation optics \cite{Pendry06, Leonhardt06, Smith06} caused a renewed 
interest for the dispersion phenomenon. Indeed, according to the causality 
principle and passivity, effective index with values below unity or negative 
requires to introduce frequency dispersion \cite{Ves68,Landau,Jackson}. 
Hence the effect of dispersion in metamaterials has been recently investigated 
in the cases of negative index, flat lens \cite{Collin10,GT10,PRL-Pen11,GM12,
PRL-Gref12} and invisibility systems \cite{Gra16}.

The original work on propagation of electromagnetic waves in homogeneous dispersive 
media led to the description of the forerunners (also known as Brillouin and Sommerfeld precursors) 
which are transients that precede the main propagating signal \cite{Bri1960}. They have been then observed later
\cite{Aav91,Jeo06,Ni06}. Since these pioneering works, the advances reported in the literature concern 
asymptotic descriptions of percursors \cite{Oug89,MS12}, the definition of energy density 
and related quantities \cite{Rup02,Dia11}, and  in dispersive and absorptive media. 
The book of Oughstun \cite{Oughstun2009} can be consulted for more complete bibliography.

A new situation is proposed to analyze the propagation of electromagnetic 
waves in dispersive media. 
The considered electromagnetic source has the sinusoidal time dependence $\sin[\om_s t]$ 
after it has been switched on at an initial time $t=0$, as introduced in 
\cite{Bri1960}, and used more recently in \cite{Collin10,GT10,GM12,Gra16}.
When such a source radiates in a dispersive homogeneous medium of 
relative permittivity $\ep(\om)$, the time dependence of the field at a 
distance $x$ from the source is given by the integral \cite{Bri1960}
\begin{equation} 
E(x,t)=\frac{1}{2\pi}\displaystyle\int_\Gamma d\om\, \dfrac{\oms}{\om^2 - \oms^2} \, 
e^{- i \om t} \; e^{i \om \sqrt{\ep(\om)} x / c_0 } \, ,
\label{intBri}
\end{equation}
where $c_0$ is the light velocity in vacuum, $\sqrt{\ep}$ is the ``index'' of the medium. 
The integration path 
$\Gamma$ is the line parallel to the real axis made of the complex frequencies
$\om = \nu + i \eta$ with imaginary part $\eta>0$. The main difficulty 
to compute this integral is the presence of the square root 
$\sqrt{\ep(\om)}$  which induces branch points and branch cuts. 

\begin{figure}[h]	
\includegraphics[width=\linewidth, keepaspectratio]{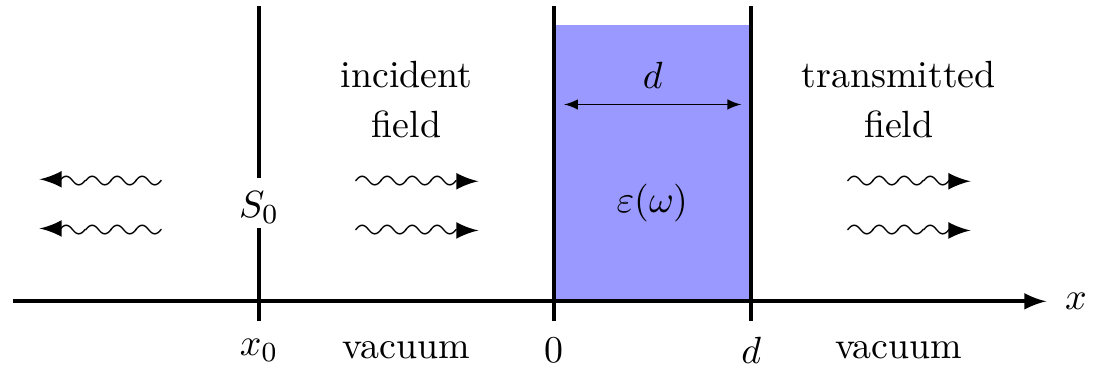}
\caption[Optional caption]{A dispersive homogeneous slab illuminated by a source $S_0$ located at a distance 
$|x_0|$.}
\label{Fig1}
\end{figure}

In this article, it is proposed to consider a one dimensional problem in the $x$-direction, 
where the source is located in vacuum at the vicinity of a homogeneous dispersive medium 
slab with thickness $d$ as illustrated in Fig. \ref{Fig1}.  The objective is to evaluate 
the transmitted field at the output.

\begin{figure} 	
\includegraphics[ width=0.7\linewidth, keepaspectratio]{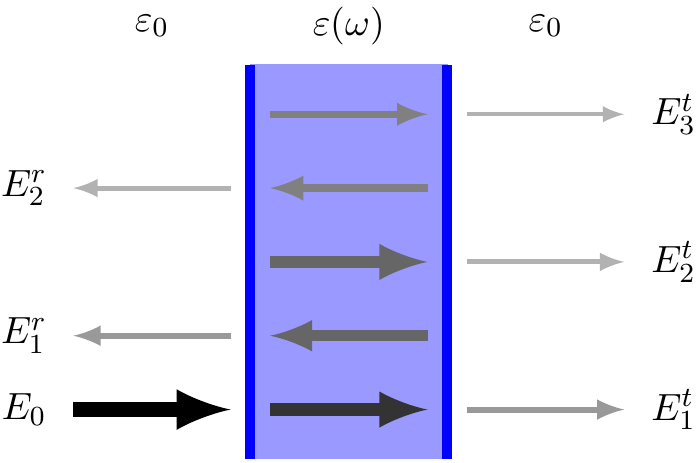}
\caption{Internal multiple reflections inside a slab of a medium with a different refractive 
index than the surroundings. Normal incidence is considered while rays are vertically shifted to 
illustrate the multiple roundtrips inside.}\label{fig:slab}
\end{figure}

The difference between the refractive indices of the slab and the surrounding media introduces 
multiple reflections inside the slab that interfere with each other depending on their relative 
phases. This can be modeled as a passive Fabry-P\'erot resonator, as shown in Fig. \ref{fig:slab}. 

In this case, the time dependence of the transmitted field at the output is 
\begin{equation} 
E(x=d,t) = \frac{1}{2\pi}\displaystyle\int_\Gamma d\om\, \dfrac{\oms}{\om^2 - \oms^2} \, 
e^{- i \om (t - x_0 /c_0)} \, T(\om) \, ,
\label{intslab}
\end{equation}
where $T(\om)$ is the transfer function of the slab. The advantage of this method follows that 
the coefficient $T(\om)$ contains no square root, therefore implying the integral expression 
free from branch points and branch cuts. Hence the value of the integral reduces to the overall 
contributions of the different poles of the equation, which can be calculated using the residue theorem. 

The purpose is to revisit the propagation of waves in dispersive media 
made by Brillouin and Sommerfeld \cite{Bri1960} in the new situation of Fig. \ref{Fig1}, 
where the dispersive medium has a finite extent and the multiple reflections effect is present 
(Fig. \ref{fig:slab}). In addition, the effect of the resonance frequency in the Drude-Lorentz 
model is exhybited. 

This article is organized as follows. After the introduction, section II describes the methodology 
used in the model leading to a general equation for the temporal response of the considered system 
of Fig. \ref{Fig1}. 
Section III deals with the non-dispersive case, which is straightforward to analyze. 
It paves the way to understand the more complicated dispersive case.
In section IV, the Drude-Lorentz model of a dispersive medium is considered. Section V 
demonstrates the temporal response of a dispersive slab, with the transient time regime including 
the precursors of Sommerfeld and Brillouin, and with the steady-state solution. 
Finally, in section VI, the case of a weakly absorbing dispersive medium is briefly approached. 
\section{Methodology \label{method}}
This section presents the method considered to address the propagation throught a dispersive homogeneous slab. 
First, the electromagnetic source is introduced and treated as a causal excitation started at a particular 
time allowing us to investigate the transient behavior of the system.  Next, the transfer 
function of the slab is given: it takes into account the internal multiple reflections, known as the Fabry-P\'erot 
resonator model. Eventually, a general equation is demonstrated for the temporal response of the dispersive 
slab that describes the transmitted field. This expression is analyzed in details in the next sections.
\subsection{The causal source} \label{sec:source}
The incident field is a monochromatic plane-wave propagating in x-direction and generated at distance $\xs$ from 
the dispersive slab, as shown in Fig. \ref{Fig1}. The source is causal: it is ``OFF'' for negative times 
and then switched ``ON'' at time $t=0$. Let $\delta$ be the Dirac function and
 $\theta(t)$ the step function: $\theta(t) = 0$ if $t<0$ and $\theta(t) = 1$ otherwise.
The considered point current source $J(x,t)$ is then given by
\begin{equation} \label{source}
J(x,t) = \dfrac{2 E_0 }{c_0 \mu_0} \, \delta(x - \xs) \, \sin[\omega_s t] \, \theta(t) \, ,
\end{equation}
where $E_0$ is a constant electric field amplitude, and $c_0 = 1 / \sqrt{\ep_0 \mu_0}$, $\ep_0$ and 
$\mu_0$ are respectively the light velocity, the permittivity and the permeability in the vacuum.
The idea is to mimic a plane wave, at the limit $t \to \infty$, with the requirement 
of causality. The spectral representation of the source components $\widehat{J}(x,\om)$
is obtained using the Laplace transform
\begin{equation} \label{eq:LT}
\widehat{f}(\om)= \int_{0}^{\infty}  dt\,  e^{i \om t} f(t) \, .
\end{equation}
Notice that the imaginary part of $\om$ must be positive to ensure the convergence of the integral. 
After this Laplace decomposition, the source becomes
\begin{equation}
\widehat{J}(x,\om) = - \dfrac{2 E_0 }{c_0 \mu_0} \, \dfrac{\omega_s}{\om^2-\om_s^2} \, \delta(x - \xs) \, .
\end{equation} 
The spectral representation of the causal monochromatic source has a maximum centered at the 
excitation frequency $\om_s$ with a nonzero broadening, in opposite to a non-causal monochromatic
 source case with a delta-function at the given frequency (pure sinusoidal wave).

\begin{figure}
\includegraphics[width=1\linewidth, keepaspectratio]{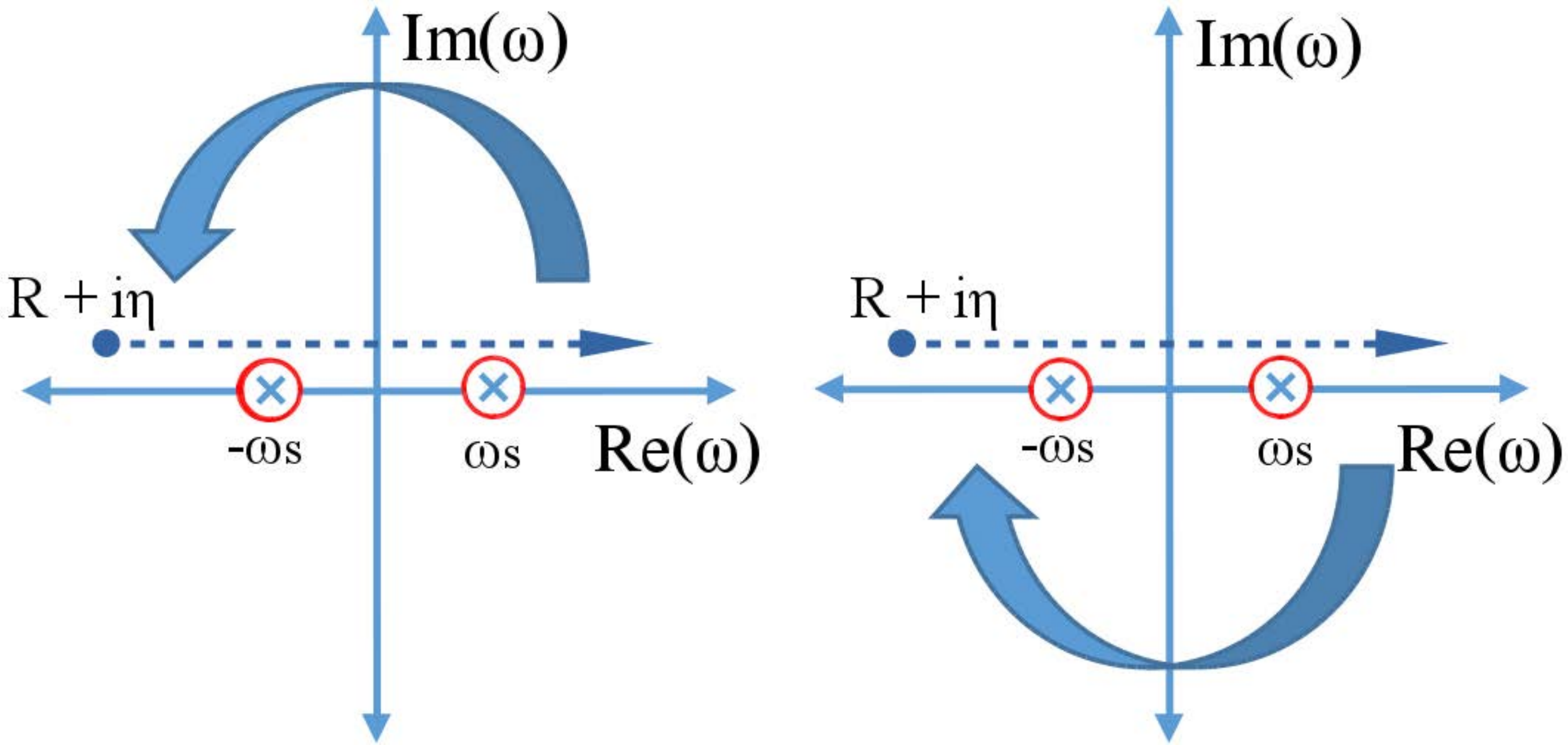} 
\caption[Optional caption]{The plane of complex frequencies $\om$ that shows the 
poles of Eq. (\ref{illumin-z}). (Left) For $t < | x - \xs | / c$, there are no poles in the 
upper half plane and the field vanishes. (Right) For $t > | x - \xs | / c$ 
the integral has a non-vanishing value.}
\label{fig:ComplexPlane}
\end{figure}

The Laplace transform is then applied to the one-dimensional Helmholtz equation to obtain the 
incident electric field radiated in vacuum:
\begin{equation}
\dfrac{d^2 \widehat{E}}{dx^2} (x,\om) + \dfrac{\om^2}{c_0^2} \, \widehat{E} (x,\om) = 
- i \om \mu_0 \widehat{J}(x,\om) \, .
\end{equation}
Taking the spatial Fourier transform into the $k$-domain, this equation becomes
\begin{equation}
\tilde{E}(k,\om) = \dfrac{2 E_0 }{c_0} \, \dfrac{\omega_s}{\om^2-\omega_s^2} \, 
\dfrac{i \om}{\om^2/c_0^2 -k^2} \, e^{- i k \xs} \, .
\end{equation}
In order to retrieve the time domain expression of the field $E(x,t)$, the inverse spatial 
Fourier transform is performed
\begin{equation}
\widehat{E}(x,\om) = E_0 \, \dfrac{\omega_s}{\om^2-\omega_s^2} \, e^{ i \om | x - \xs | / c_0} \, .
\end{equation}
Here, the absolute value preserves the direction of propagation to be 
as outgoing waves. Then, applying the inverse Laplace transform leads to the final 
expression of the illuminating field, normalized to the input amplitude $E_0$:
\begin{equation}\label{illumin-z}
\begin{array}{ll}
E(x,t) & = \dfrac{1}{2\pi}\; \displaystyle{\lim_{\Gamma \to 0}} \displaystyle\int_{\Gamma} d\om \, e^{-i \om t } \,  
\dfrac{\omega_s}{\om^2-\omega_s^2} \, e^{ i \om | x - \xs | / c_0} \, .
\end{array}
\end{equation}
The integration is performed along the line $\Gamma = \mathbb{R}+i\eta$ in the complex frequency 
domain in order to prevent the divergence of the integration, as shown in Fig. \ref{fig:ComplexPlane}. 
This satisfies the causality principle that requires the existence of the field only for 
 $t > | x - \xs | / c_0$ where the integral has a non-zero value due to the presence of poles 
below the integration line on the real axis. 

\subsection{The resonator model of the slab} \label{sec:slab}
A plane wave encountring an interface between two different media is subject to reflection. Assuming 
the surrounding medium of the slab is vacuum, the portion of the reflected field from a single interface, 
i.e. the reflection coefficient $\rho$, is given by
\begin{equation}\label{eq:r}
\rho(\om) = \dfrac{1 - \sqrt{\ep(\om)}}{1 + \sqrt{\ep(\om)}} \, .
\end{equation}
The wave propagating through the slab experiences multiple reflections which create other copies 
of the main signal with amplitudes depending on the relative refractive index of the slab to the surrounding 
medium. These multiple copies interfere according to the phase difference between them, which is determined 
by the propagation length inside the slab and the frequency of the excitation. Subsequently, the slab can be 
modeled as a frequency-dependent element, i.e. a resonator.

\begin{figure}
\includegraphics[ width=1\linewidth, keepaspectratio]{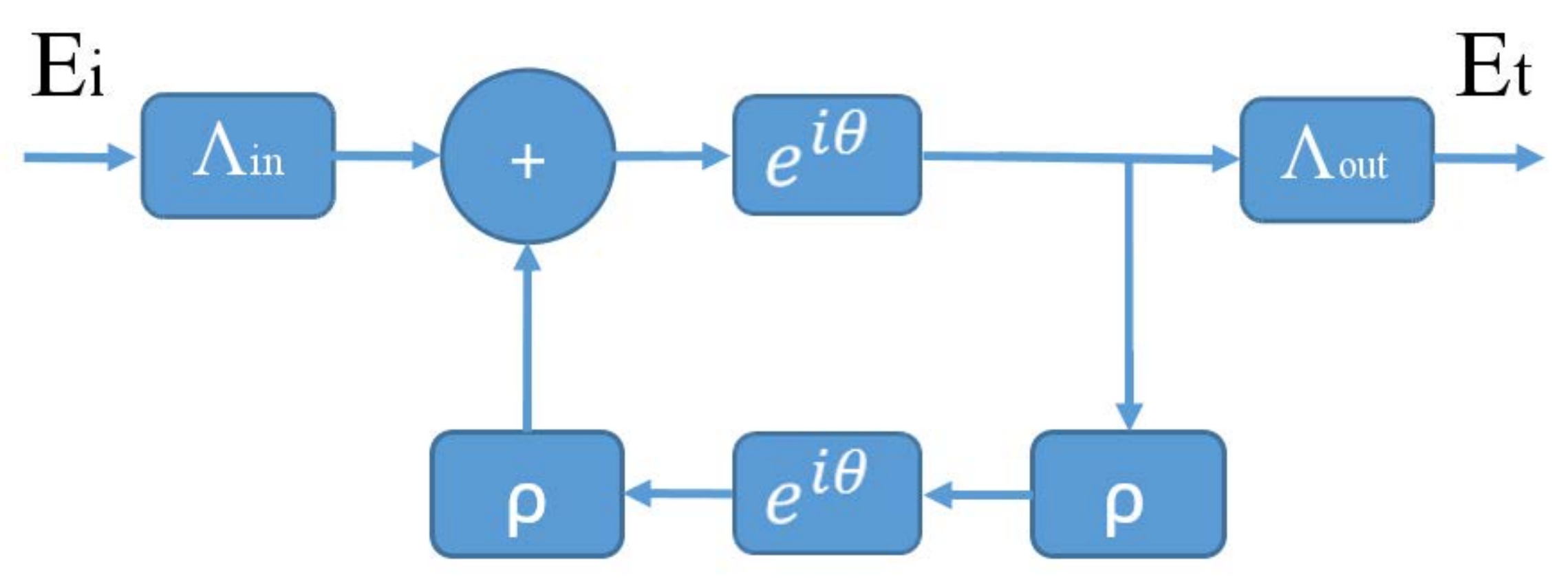}
\caption[Optional caption]{The resonator block diagram that resembles the propagation inside 
a slab taking into account the internal multiple reflections.} 
\label{fig:resonator}
\end{figure}

A block diagram of the resonator is shown in Fig. \ref{fig:resonator}, where $\theta$ 
is the 
propagation phase during a single trip and $\Lambda$ is the transmission coefficient of a single interface, 
related to the reflection coeffcient throught $\Lambda = 1 - \rho$. The loop represents the round-trip 
due to the internal reflections between the slab interfaces.

Normalized quantities are used to describe the system in a general way, with the frequency and the time 
normalized with respect to the slab thickness $d$ and the ligth velocity in vacuum:
\begin{equation}
\hom = \dfrac{\om d}{c_0} \longrightarrow \om \, , \quad
\hat{t}= \frac{c_0 \, t}{d} \longrightarrow  t \, .
\label{hom}
\end{equation}
The circumflex is omitted in the rest of this letter.

The slab transfer function $T(\om)$ can be obtained by applying the feedback theory \cite{sedra}
\begin{equation} \label{eqSlabTF}
T(\om) = \dfrac{[ 1 - \rho(\om)^2 ] \, e^{i \om \sqrt{\ep(\om)} }}{1-\rho(\om)^2 \, 
e^{2 i \om \sqrt{\ep(\om)}}} \, ,
\end{equation}
which provides the same result as the electromagnetic calculation.
It is important to remark that this transfer function can be written as
\begin{equation}\label{eqT}
\dfrac{1}{T(\om)} = \cos[\om \sqrt{\ep(\om)}] + i \, \dfrac{1 + \ep(\om)}{2} \, 
\dfrac{\sin[\om \sqrt{\ep(\om)}]}{\sqrt{\ep(\om)}} \, .
\end{equation}
Indeed, this expression shows that $T(\om)$ is an even function of $\sqrt{\epsilon(\om)}$. 
This implies the absence of square roots of permittivity in $T(\om)$, hence the absence 
of branch cuts in  Eq. (\ref{intslab}). This is an advantage for the resonator model in 
comparison with the method generally used \cite{Bri1960}, since dealing with branch cuts in Eq. (\ref{intBri}) 
is a challenging step in the study of propagation inside a dispersive medium \citep{Bri1960}. 

\subsection{Temporal response equation} \label{sec:tempeq}
Replacing the slab transfer function by its expression (\ref{eqSlabTF}) in the integral 
(\ref{intslab}), and taking the source location at the edge of the slab $x_0=0$, the 
transmitted field at the output of the slab ($x=d$) is
\begin{equation}\label{eq:ResponseInt}
E_d(t) =\frac{1}{2\pi} \int{d\omega\;e^{-i \omega t } \; \frac{\omega_s}{\omega^2-\omega_s^2} 
\;\frac{ [1-\rho^2(\om)]\; e^{i \om \sqrt{\epsilon(\om)}}}{1- \rho^2(\om)\;e^{2i \om \sqrt{\epsilon(\om)}}} } \, ,
\end{equation}
which is the general equation to describe the response of a slab of a dispersive medium 
to a causal excitation. Since no branch cut exists, the integral reduces to the 
contributions of discrete poles given by the residue theorem. The poles of the integral are 
defined by
\begin{equation} \label{eq:poleseq} 
\pm \om_s \, , \quad \om_q: \rho(\omega_q) = \pm e^{i\; \sqrt{\epsilon(\omega_q)} \omega_q} \, .
\end{equation}
Here, the couple $\pm\om_s$ contains the source poles providing to the steady state part of 
the solution defined as ($t\to\infty$), 
while the infinite set of $\om_q$ contains the poles of the slab transfer function $T(\om)$ 
corresponding to the transient part of the solution. Notice that the reason of the existence 
of the transient behavior is the causality of the source.

The solution of the output field can be written as, 
\begin{equation} \label{twoparts} 
\begin{array}{rl}
E_d(t) = & \theta(t -  \tau_0)\;  \, E_{\text{stst}}(t) \\[4mm]
+ & \theta(t - \tau_0)\; \, E_{\text{trans}}(t) \, , 
\end{array}
\end{equation}
where $\tau_0= 1$ is the single-trip normalized time needed for the front of the wave 
(fastest part that experiences a unity permittivity) to reach from the input side of the slab 
to the output side. This preserves causality since no signal can reach from one side to the 
other faster than $c_0$.

The steady state part can be calculated using that $T(-\om)$ is the complex conjugated of $T(\om)$. 
It gives a causal function oscillating at the excitation frequency $\om_s$, multiplied by a scaling factor 
depending on the slab characteristics:
\begin{equation} \label{eq:stst} 
E_{\text{stst}}(t)= - \text{Im}
\big[ e^{- i \oms t }\; T(\oms) \big] ,
\end{equation}
where ``Im'' means the imaginary part. 
As to the transient part, it is given by the residues of the set of poles $\om_q$ of the 
transmission coefficient $T(\om)$. Since $T(\om)$ is not polynomial, 
then the transient part can expressed as
\begin{equation} \label{eq:MainTransient}
E_{\text{trans}}(t)= i  \;\sum_{q} e^{-i \om_q t } \; \frac{\omega_s}{\om_q^2-\omega_s^2}\; 
\left[ \dfrac{\partial T^{-1}} {\partial \om} \, (\om_q) \right]^{-1}\, , 
\end{equation} 
where $T^{-1}(\om)$ is considered as the denominator of the transfer function $T(\om)$. The 
expression (\ref{eqT}) leads to 
\begin{equation} \label{eq:Qdiff}
\begin{array}{rl}
\dfrac{\partial T^{-1}} {\partial \om} \, (\om_q) = & 
i \, \dfrac{1 + \epsilon(\om_q)}{2} \left[ 1 + \dfrac{\om_q}{2 \epsilon(\om_q)} 
\dfrac{\partial \ep}{\partial \omega} (\om_q) \right] \\[4mm]
& \times \cos[\om_q \sqrt{\epsilon(\om_q)}] \\[4mm]
- & \left[ \epsilon(\om_q) + \dfrac{2 \om_q \epsilon(\om_q) + i (\epsilon(\om_q) - 1)}{4 \epsilon(\om_q)}
\dfrac{\partial \ep}{\partial \omega} (\om_q) \right] \\[4mm]
& \times 
\dfrac{\sin[\om_q \sqrt{\epsilon(\om_q)}]}{\sqrt{\epsilon(\om_q)}}

\end{array}
\end{equation}
Each term in the sum (\ref{eq:MainTransient}) represents the contribution of a pole $q$ 
to the total transient response of the system and has a scaling coeffcient depending on the 
distance to the source pole. 

It is stressed that this method presents the crucial advantage of the resonator model,
when compared with the usual integral Eq. (\ref{intBri}), as as it preserves to any branch cut
and hence reduces to the use of the residue theorem to obtain the expression of the field propagating 
in a dispersive medium.
\section{Non dispersive  slab\label{method}}

\begin{figure} 	
 \includegraphics[width=0.75\linewidth, keepaspectratio]{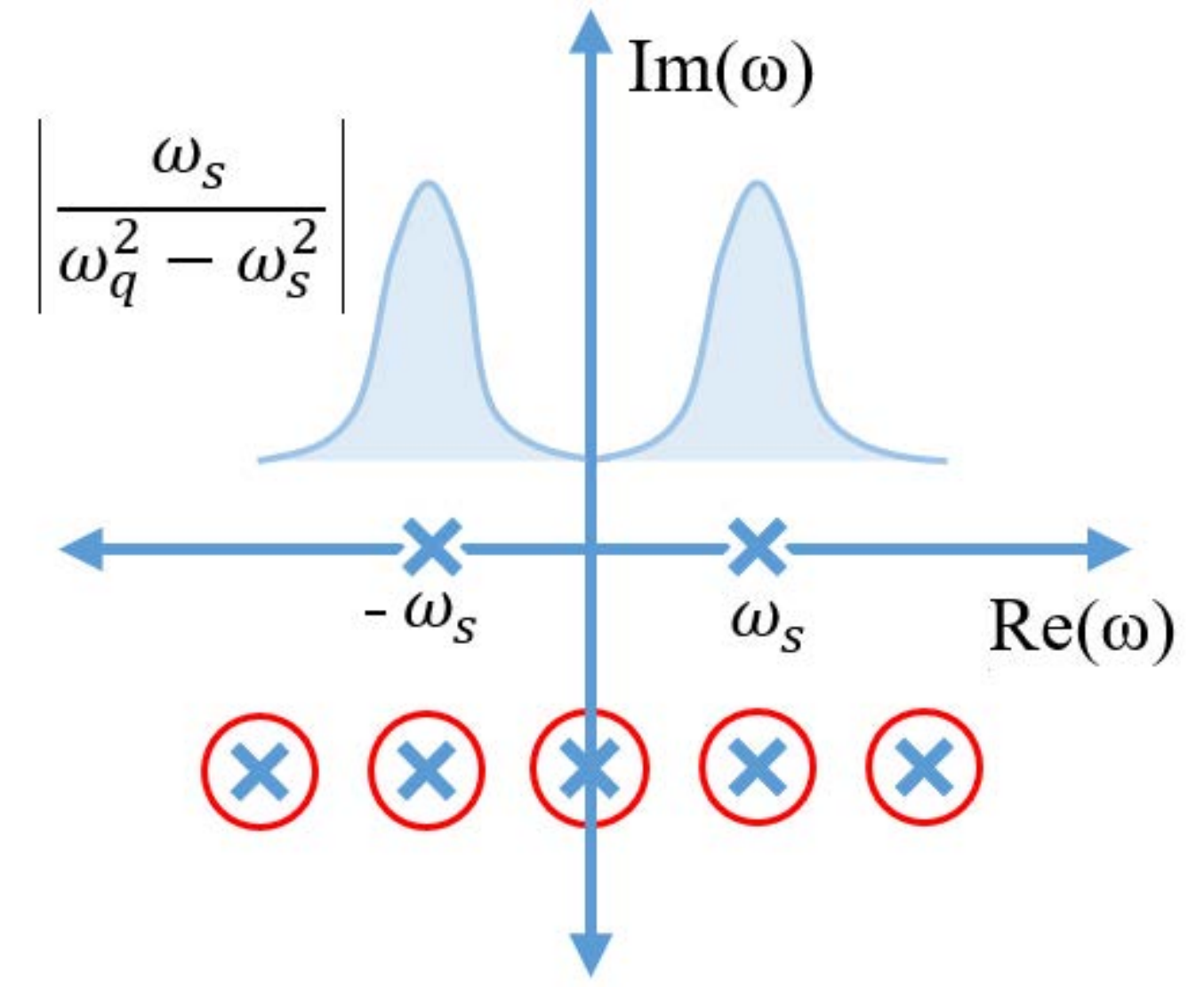}
 \caption[Optional caption]{ Equally distant poles of a nondispersive slab with the same imaginary part.
  The curve sketches the relative contributions of the slab poles as a function of the distance to the source poles.}
 \label{fig:nondispoles}
 \end{figure}

A non dispersive dielectric slab with permittivity fixed to $\ep(\om)=\ep_s$ has 
source poles $\pm\om_s$ and the slab poles
\begin{equation}  \label{eq:nondisppoles}
\om_q = \frac{q \pi}{\sqrt{\ep_s}} - i \frac{\ln (1/\rho_s^2)}{2\sqrt{\ep_s}},
\end{equation}  
where $q$ is an integer. Figure \ref{fig:nondispoles} shows both types of poles in 
the frequency domain, where it can be checked that $\om_{-q}= - \overline{\om_q}$. 
The imaginary part represents the losses 
of the resonator (reflection losses and, if it is present, absorption). Assuming a lossless medium, 
the wave propagating inside suffers only 
from the reflection (outcoupling) loss at the interfaces. The more outcoupling is, the 
less round-trips the wave stays inside, the shorter the transient regime is.

The single-trip time value determines the initial time at which the transmitted field starts to appear
 at the output,
\begin{equation} \label{eq:Taues}
\tau_s= \tau_0\; \sqrt{\ep_s} \, .
\end{equation}
Since the contribution of each pole depends on the distance to the source pole, it is 
assumed that the closest two slab poles ($\pm q'$) to the source poles are dominant in the 
transient summation as $|\om_{\pm q'}| \approx |\om_s|$ and $\text{Re}({\om_{\pm q'}})=\om_s$, 
leading to
\begin{equation} \label{eq:nondisresp} 
\begin{array}{ll}
E_{\text{d}}(t) \approx & - \text{Im} \big[ e^{-i\om_s t}\; T({\oms}) \big] \; \theta(t -  \tau_s) \\[4mm]
& + \text{Im} \big[ 2\,e^{-i \oms t } \, e^{-t/\tau_{\text{tr}}} F(\oms) \big] \; 
\theta(t -  \tau_s),
\end{array}
\end{equation}
where the first part represents the steady state solution, and the other part shows the transient part, 
given that,
\begin{equation}
F(\oms) = \frac{ (1-\rho_s^2)\; e^{i \oms \sqrt{\epsilon_s} } } {2\; \ln(1/\rho_s^2)} \, , \quad 
\tau_{\text{tr}} = \dfrac{2\sqrt{\ep_s}}{\ln (1/\rho_s^2)} \, .
\end{equation}
The time-constant $\tau_{\text{tr}}$ which characterizes the decay of the transient region is 
the reciprocal of the imaginary part of the poles given in Eq. (\ref{eq:nondisppoles}).

\begin{figure} 	
 \includegraphics[width=0.85\linewidth, keepaspectratio]{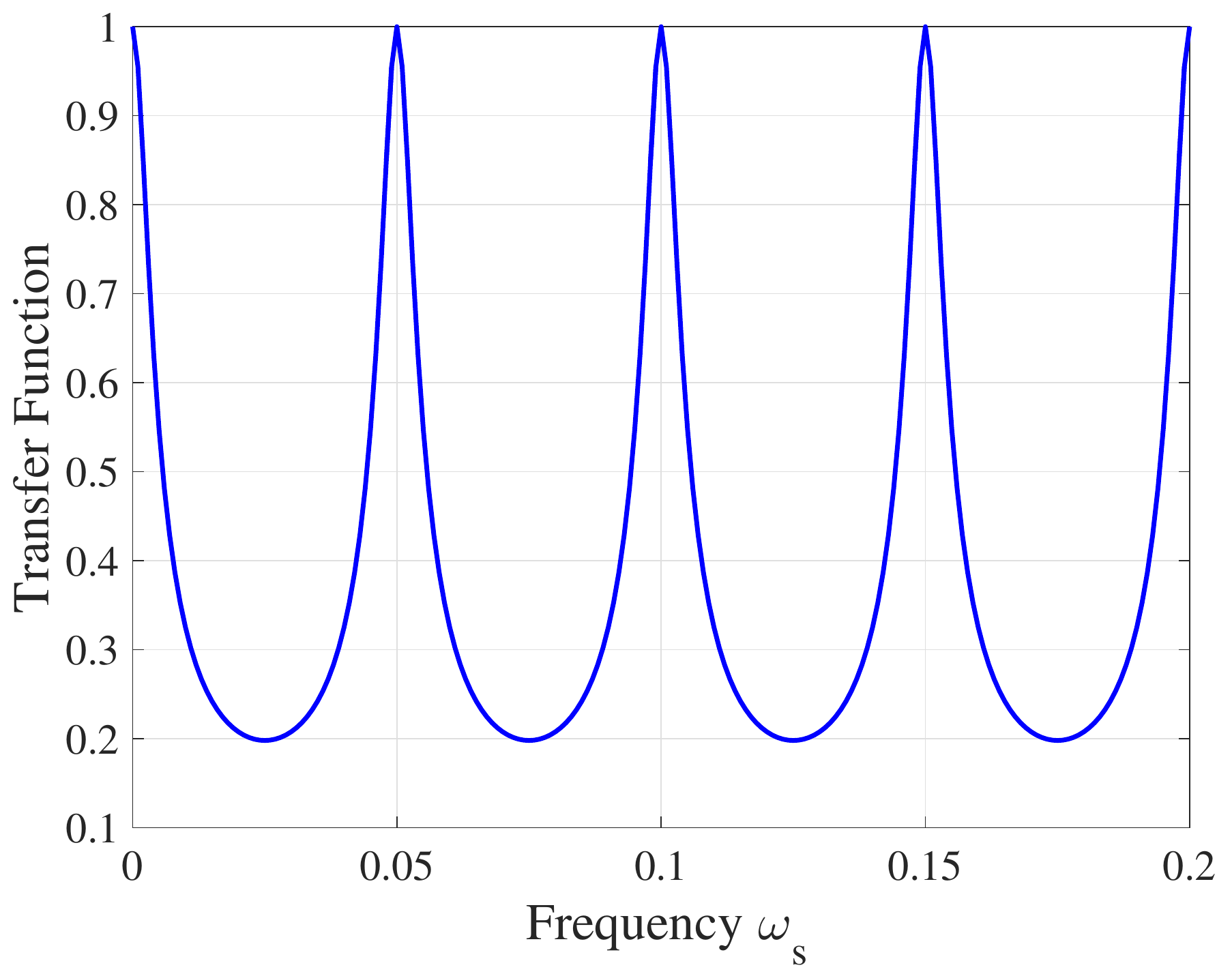}
 \caption[Optional caption]{
The transfer 
 function of a non dispersive  slab with a dielectric constant $\eps=100$. }
 \label{fig:nondispws}
 \end{figure}

\begin{figure} 	
 \includegraphics[width=0.85\linewidth, keepaspectratio]{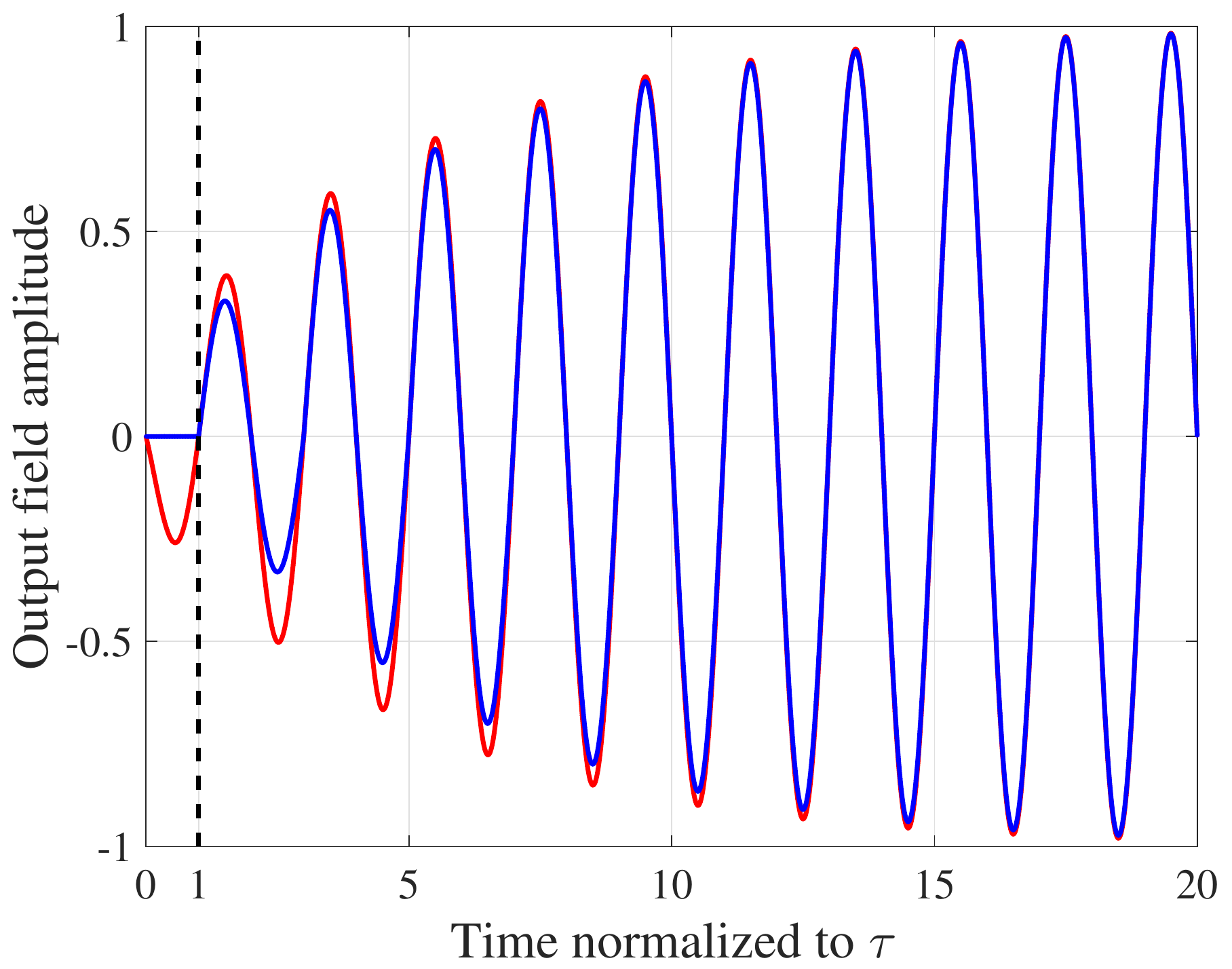}
 \caption[Optional caption]{Temporal response 
(time normalized to $\tau_s$)
 of the non dispersive slab for an excitation 
 frequency $\om_s=0.05 \times 2 \pi$ showing a maximum transmission. 
 The red curve shows the approximated expression.
 [The step function $\theta(t -  \tau_s)$ has not been implemented in 
 this curve, which leads to a non vanishing contribution from $\tau = 0$ 
 to $\tau = \tau_s$.] 
}
 \label{fig:nondisw005}
 \end{figure}

When the source frequency matches the real part of one of the poles, the transmission 
is maximized (unity for a symmetric resonator). The interference between the multiple 
reflections inside is constructive in this case as the phase difference is always a 
multiple of $2\pi$.  The frequencies at which minimum transmission occurs called 
anti-resonant frequencies and the minimum transmission value depends on the single interface 
reflectivity,
\begin{equation} \label{eq:Tmin}
T_{\text{min}} =  \frac{1-\rho^2}{1+\rho^2}.
\end{equation}
For a test case of a dielectric constant $\eps = 100$, Fig. \ref{fig:nondispws} shows the 
slab transfer function modulus $|T(\om_s)|$ as the function of the excitation frequency, which 
demonstrates the effect of the internal multiple reflections. Figure \ref{fig:nondisw005} 
shows the temporal behavior of a dielectric slab for an excitation frequency 
corresponding to a maximum transmission at the steady state regime. 
The blue curve in the figure shows the temporal response using Eq. (\ref{twoparts}) 
that includes the contributions of all poles, while the red curve is using the 
approximated Eq. (\ref{eq:nondisresp}) that only includes the two nearest poles 
to the sources poles. 


\section{Dispersive medium Model} \label{sec:dismat}
In this article, the Drude-Lorentz model is used for the dispersive medium \cite{Ckittle}, 
similarly to Brillouin and Sommerfeld \cite{Bri1960}. The frequency dependency 
of the permittivity is then given by
\begin{equation}
\ep(\om) = 1 - \frac{\Om^2}{\om^2-\omr^2+i\om \gamma}  \, 
\label{epDL}
\end{equation}
where $\Om$ is a constant of the medium (related to the electron density \cite{Jackson}), 
$\omr$ is the resonance frequency of the dispersive medium, and $\gamma$ is 
the absorption constant. 

We are interested in the case of lossless dispersive medium, 
i.e. $\gamma = 0$. It is stressed that modeling the lossless case 
should be addressed by taking the limit of Eq. (\ref{epDL}) as $\gamma \downarrow 0$. 
According to causality principle and the Kramers Kronig relations \cite{Jackson}, 
the dispersion leads to an imaginary part in the permittivity. Indeed, in the limit 
$\gamma \downarrow 0$, the imaginary part of the permittivty turns to be a Dirac function 
at $\pm \omr$. Nevertheless, this Dirac contribution can be ignored because the transmission 
of the input interface of the slab vanishes as $\ep \to \infty$ at $\pm \omr$. We will show 
later that setting the absorption term to 0 gives the same results as taking the limit, 
see section  \ref{sec:NumValid}. In this way, we can confidently use the Lorentz model 
for the lossless case by setting $\gamma$ to zero. 
Under this condition, the plasma frequency  $\omp$, defined by the permittivity set to zero, 
is 
\begin{equation}
\omp^2 = \omr^2+\Om^2. 
\end{equation}

\begin{figure}
\includegraphics[width=0.85\linewidth, keepaspectratio]{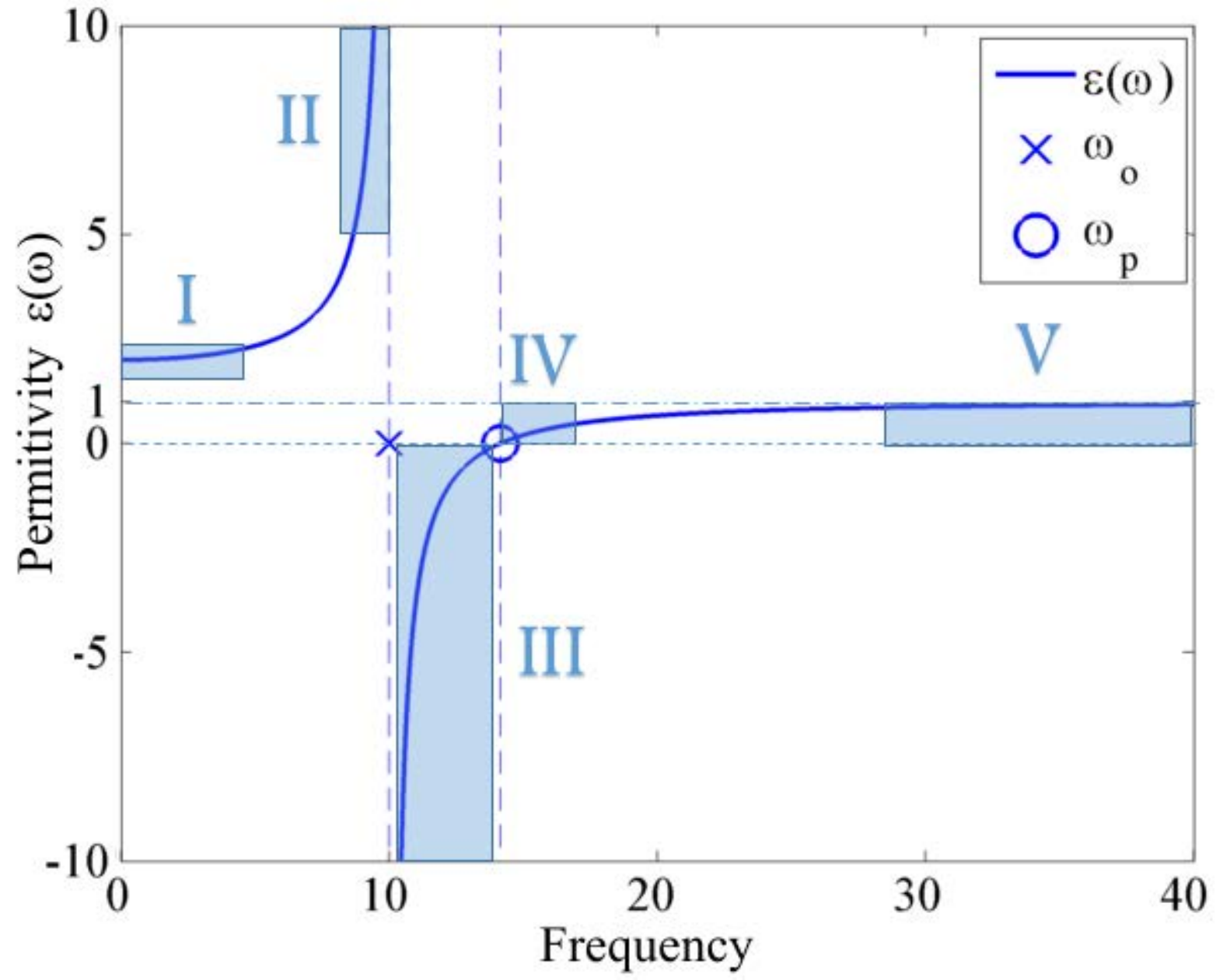}
\caption[Optional caption]{The Drude-Lorentz model of the permittivity of a dispersive medium with $\omr=\Om=10$. }
\label{fig:epsilon} 
\end{figure}

Figure \ref{fig:epsilon} shows the  permittivity of a dispersive medium with $\omr=\Om=10$ 
and $\om_p=10\sqrt{2}$ as a test case. Determining exactely the poles of the dispersive slab 
analytically is unattainable. However, it is possible to obtain estimates if the problem is 
analyzed in different frequency zones where one can approximate the permittivity. 
Five important zones, that have distinct features where an analytical solution 
can be found, are considered separately. As to the poles in a region in between these zones, 
they can be interpolated from the two adjacent ones. 

It is relevent to investigate the group velocity of a wave in a dispersive 
medium \cite{Bri1960}, which corresponds to the velocity of the pulse envelope:
\begin{equation}
\dfrac{v_g(\om)}{c_0} = \frac{1}{n_g(\om)} = \dfrac{1 / n(\om)}{1+ \frac{\om}{n(\om)} \frac{dn(\om)}{d\om}}.
\end{equation}
where $n_g$ is the group refractive index of the medium that defines the propagation of a 
narrow-band pulse in a dispersive medium. It can be associated with the velocity of the energy 
propagation of the pulse which cannot exceed the speed of light: whenever $\ep(\om) > 0$, 
which corresponds to a propagating solution with a real refractive index, $v_g(\om)$ is always below $c_0$.
 
\begin{figure}
\includegraphics[width=0.85\linewidth, keepaspectratio]{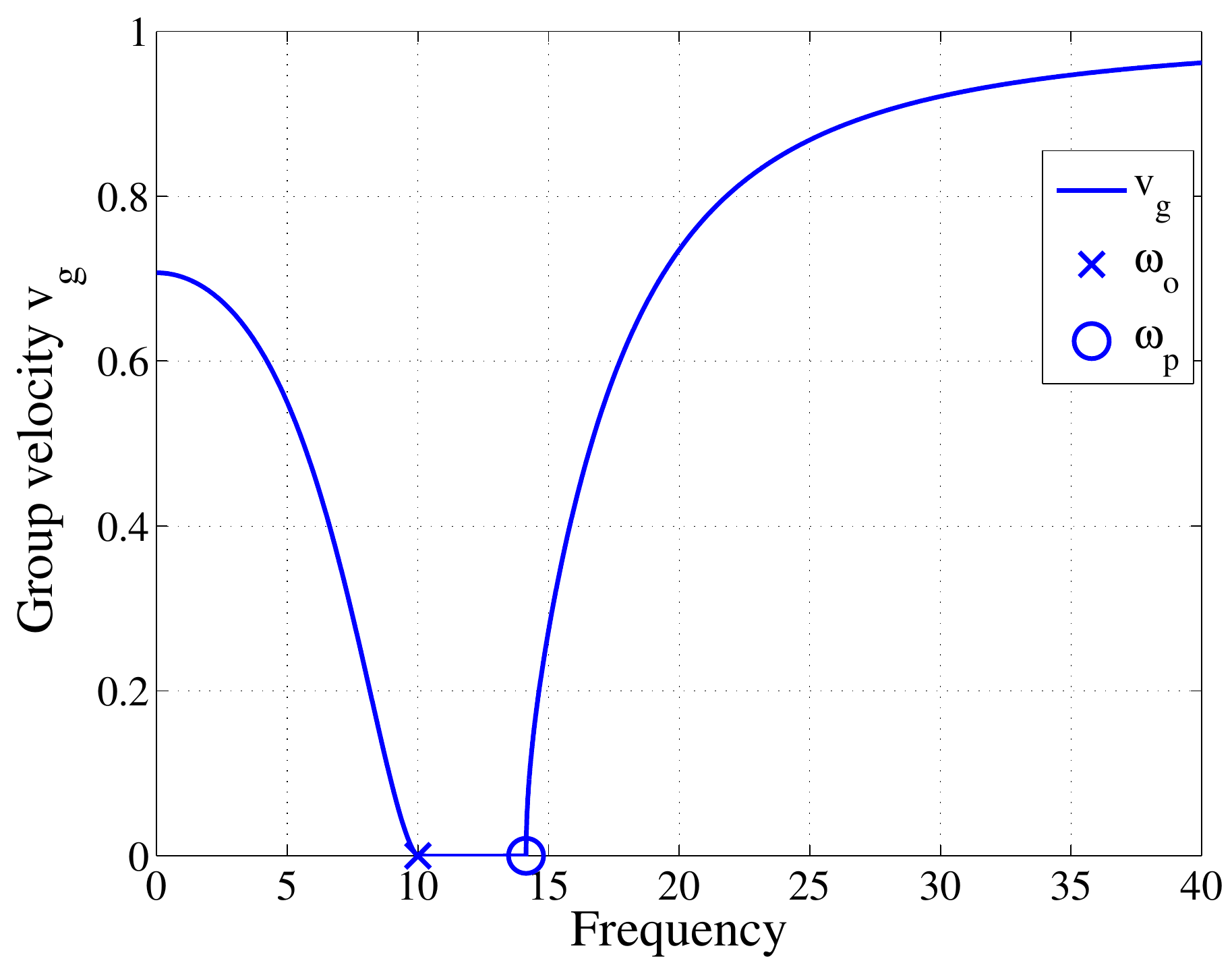}
\caption[Optional caption]{The group velocity of the dispersive medium with $\omr=\Om=10$.} 
\label{fig:vg} 
\end{figure}

Figure \ref{fig:vg} shows the group velocity as a function of the frequency for the same dispersive 
medium with $\omr=\Om=10$. This information can be used to visualize the temporal response 
of a dispersive system.  The single-trip time for a narrow-band pulse propagating in a dispersive medium is then
 \begin{equation}  \label{eq:transtime}
 \tau_g = \tau_o \; n_g,
 \end{equation}
where $n_g$ is defined for the central frequency of the pulse.
\subsection{Zones of Dispersive medium} \label{sec:poles}
The analytical estimates of the poles of the dispersive slab are provided for the different 
zones represented on Fig. \ref{fig:epsilon}. The detailed calculations are reported in the appendix.
\subsubsection{{Low-frequency zone}: $\om \ll \omr$ and $\ep \simeq \eps > 1$}
In this zone, the permittivity can be approximated as
\begin{equation} \label{eq:zone1eps}
\ep(\om)  \underset{\om \ll \omr}{\approx}  \eps + (\eps - 1 ) \, \dfrac{\om^2}{\omr^2} \, ,
\end{equation} 
where $\eps = 1 + \Om^2 / \omr^2$ is the static permittivity at $\om = 0 $.
The medium can be considered non dispersive as the group velocity is almost 
constant in this zone. The obtained expression of the poles is,
\begin{equation}
\label{eq:polesnearZ}
\omega_q = \frac{q\pi }{\sqrt{\ep_s}} - i \dfrac{\ln \big| 1/\rho_{\om_s} \big|}{\sqrt{\eps}}
+ \dfrac{q^2 \pi^2}{8 \omr^2 \eps \sqrt{\eps}} \, ,
\end{equation}
and remains valid as long as $q \pi \ll \sqrt{\ep_s} \omr$.
\subsubsection{Near-resonance zone: $\om <\approx \omr$ and  $\ep \to +\infty$} 
In this case, the permittivity $\ep \to +\infty$ and it can be approached by
\begin{equation} \label{epr}
\ep(\om)  \underset{\om \to \omr}{\approx}  \dfrac{\Om^2}{2 \omr} \, \dfrac{1}{\omr - \om} \, . 
\end{equation} 
This zone is highly dispersive while the group velocity tends to zero. 
The poles are found using an iterative method (see appendix):
\begin{equation}\ \label{eq:NearPoles}
\omega_{q} = \omr -  \frac{\omr  \Omega^2 }  {2  q^2 \pi^2 } \left[ 1 +i \frac{4\omr}{q^2\pi^2} \right].
\end{equation}
This expression is limited to $q \gg  (\omr,\Omega)$ and no pole exists for $\om >\approx \omr$.
\subsubsection{Damping zone: $\omr<\om<\omp$ and $\ep < 0$}
Here, the permittivity $\ep (\om)$ is negative and the index purely imaginary. 
This leads to a decay of the signal inside the medium, so there is no wave propagation 
and no poles in the transfer function. 
This zone can be thus ignored for the calculation of the time dependent field.
\subsubsection{Near-plasma zone: $\om >\approx \om_p$ and $\ep >\simeq 0$} 
The dielectric becomes transparent again, its expression can be approached by
\begin{equation} 
\ep(\om) \underset{\om \to \omp}{\approx} 
\dfrac{\om^2 - \omp^2}{\omp^2 - \omr^2},
\label{epp}
\end{equation}
and the set of poles is given by
 \begin{equation}\label{eq:PlasPoles}
\omega_{q}=\omega_p \; \left[ 1+ \frac{q^2 \pi^2 \Omega^2}{8 \omega_p^4}  \;(1 - 4 i / \omega_p) \right].
\end{equation}  
The limit of this expression is $q \ll \frac{\sqrt{8} \omega_p^2}{ \Omega}$.
\subsubsection{{High-frequency zone}:  $\om \gg \omr$ and $\ep \simeq 1$} 
The medium cannot follow the excitation of the source and behaves like the vacuum, 
and $v_g \to c_0$. The permittivity is close to unity and the poles are given by
\begin{equation}\label{eq:zone5poles}
\omega_{q} = {q\pi} - i \ln{\frac{q^2 \pi^2 -\omr^2}{\Omega^2/2}} \, .
\end{equation} 
This expression for high frequency poles remains valid as 
long as $\pi^2 q^2 \gg \omr^2 + \Omega^2 /2 $.

\subsection{Numerical Validation and discussion} \label{sec:NumValid}

\begin{figure}	
  \includegraphics[width=0.9\linewidth, keepaspectratio]{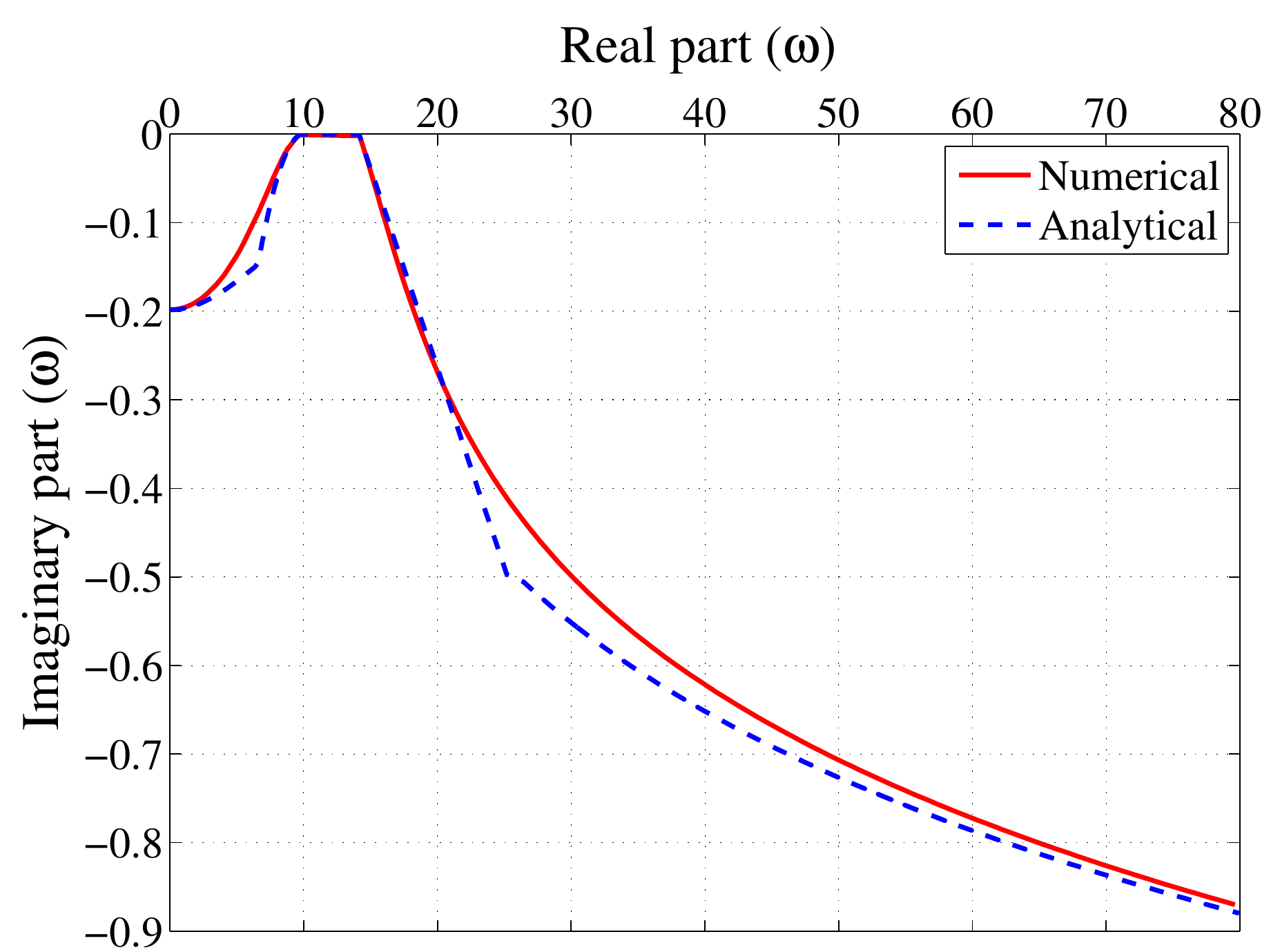}
 \caption[Optional caption]{Analytical vs. Numerical results for the poles of the test case. }
 \label{fig:allzones}
 \end{figure}

\begin{figure} 	
  \includegraphics[width=0.9\linewidth, keepaspectratio]{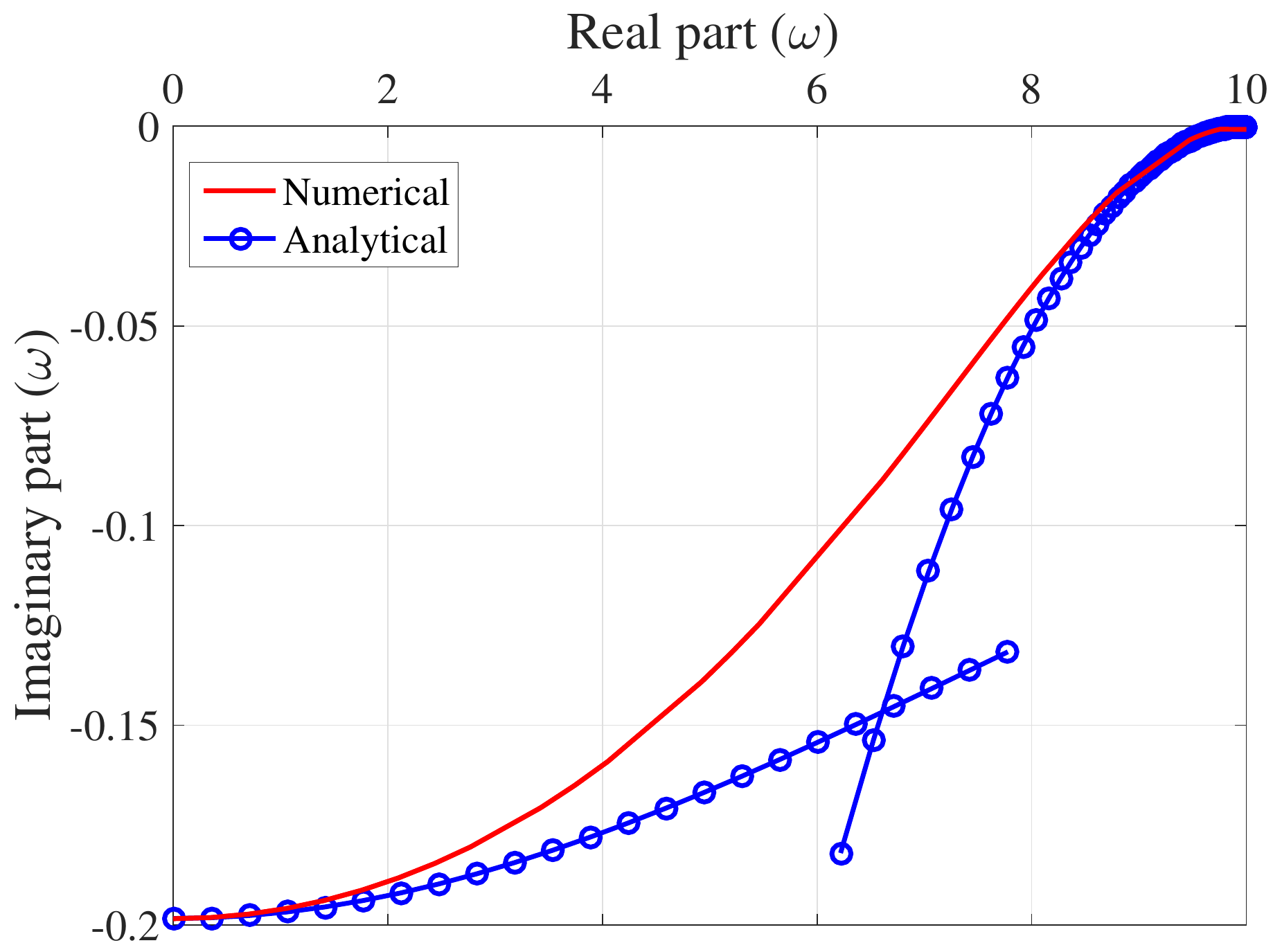}
 \caption[Optional caption]{The poles below resonance of the test case.}
 \label{fig:PolesDCRes}
 \end{figure}

Using Muller's method \cite{Mullernumerical}, one can check numerically the derived expressions 
of the poles of a test case of a dispersive slab with $\omr=\Omega=10$. Figure 
\ref{fig:allzones} shows the comparison between the analytical and the numerical results for the poles.
Figure \ref{fig:PolesDCRes} illustrates the region below resonance and show an example of how we 
can asymptotically use the derived expressions to evaluate the poles of a dispersive slab 
anywhere in the frequency domain. 

For a dispersive system, the imaginary part of the poles are frequency dependent in contrast 
to the non dispersive case. This imaginary part defines the transient time of the system 
since it represents its decay: it is related to the losses due to refections and absorption. 
For near-resonance zone, approaching $\omr$ leads to a unity 
reflectivity [since $\rho \to -1$, see Eq. (\ref{eq:r})]. Thus, the imaginary part of the 
poles tends to zero. This means that the wave at this frequency undergoes low losses and therefore it 
stays longer inside the resonator. Similarly, approaching $\omega_p$ leads to the imaginary part
of the poles tending to zero. As the frequency increases (zone 5), the imaginary part increases as well 
and $\rho \to 0$ as $\ep \to 1$ meaning that most of the wave will escape the 
resonator without suffering from a reflection. All these cases are discussed 
in detail in the next section. 

\begin{figure}	
  \includegraphics[width=0.9\linewidth]{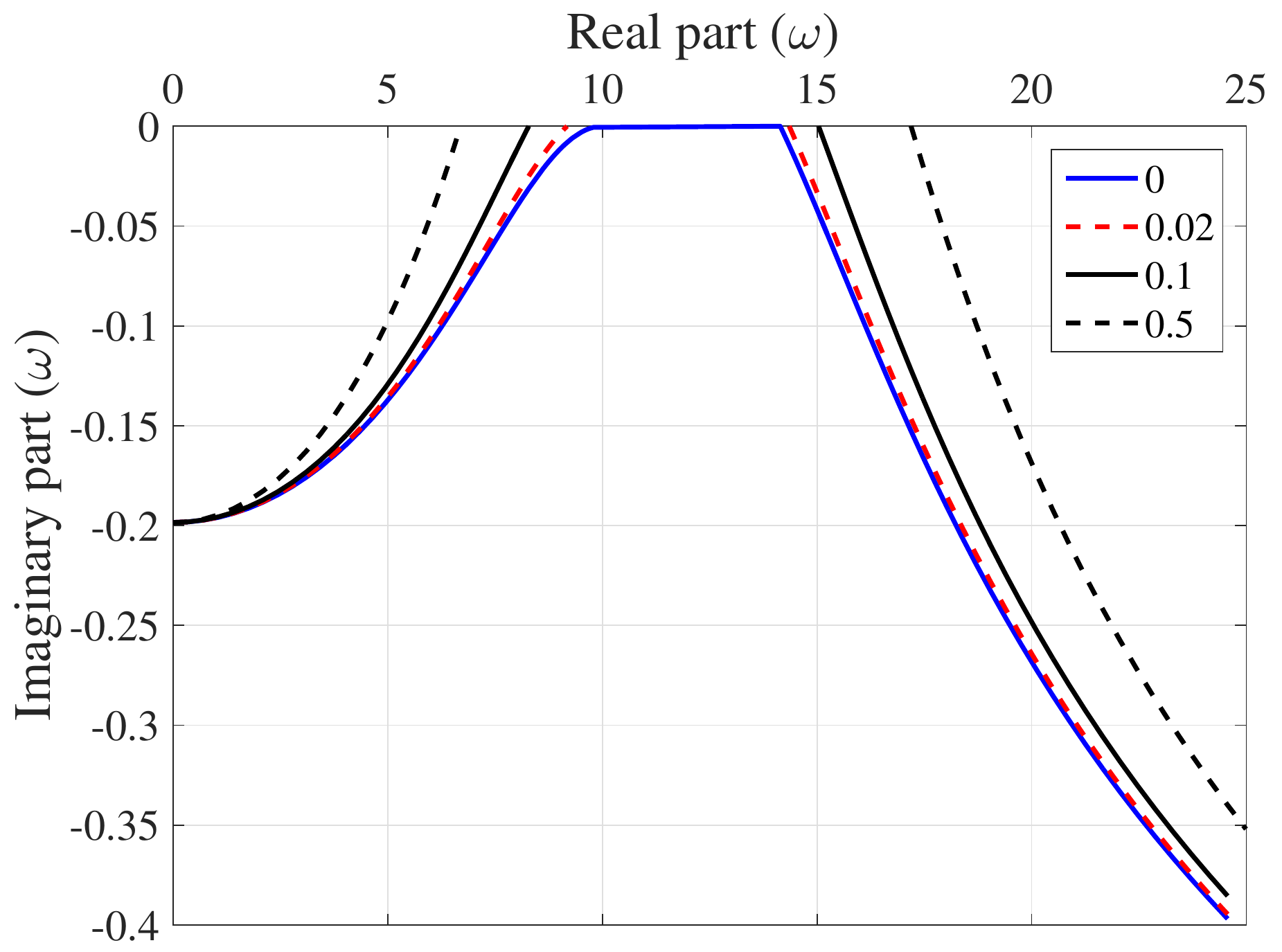}
 \caption[Optional caption]{The validation of Lorentz model of dispersive medium for the lossless case. }
 \label{fig:LorentzCONV}
 \end{figure}

Fig. \ref{fig:LorentzCONV} shows the validity of the Drude-Lorentz model to study the
lossless dispersive medium excited by a causal source, as discussed in the beginning of 
this section. Taking the limit of the absorption term $\gamma \to 0$ gives the same 
results as putting $\gamma = 0$ in the Drude-Lorentz model (blue curve on Fig. \ref{fig:LorentzCONV}).
\section{The dispersive slab response } \label{sec:response}
We turn now to the analyzis of the transmitted field throught the dispersive slab.

According to Fig. \ref{fig:vg}, a causal pulse (spectral broadened) is expected to 
split into different temporal regions while propagating in a dispersive medium as each 
frequency zone of the pulse has its own group velocity. This section represents the response of a 
dispersive slab that gives the transmitted field using Eqs. (\ref{twoparts}-\ref{eq:MainTransient}).

We represent the temporal solutions of the different frequency zones, in a chronological 
order according to the value of the group velocity. The poles of zone 5 having group velocity 
close to unity appears first at the output: the corresponding contribution in the field 
is known as the Sommerfeld precursor (or forerunner). Then, it is followed 
by the Brillioun precursors which is the response of the poles of zone 1. The poles of zone 2 
and 4 appear later as they have smaller group velocities. The transient and steady state 
solutions should add together so that the main signal starts to appear at the output after 
a certain time that depends on the group velocity value at the source frequency. 

We consider the same test case of $\omr = \Om = 10$, excited by a source frequency 
$\om_s = 16 $ that falls in the plasma zone. For this case $n_g(\om \approx 0)=\sqrt{2}$ 
and $n_g(\om_s)=2.36$. These values determine the beginning of the Brillouin 
precursor and the main pulse, respectively.
\subsection{Sommerfeld precursor}

\begin{figure} 
  \includegraphics[width=0.9\linewidth]{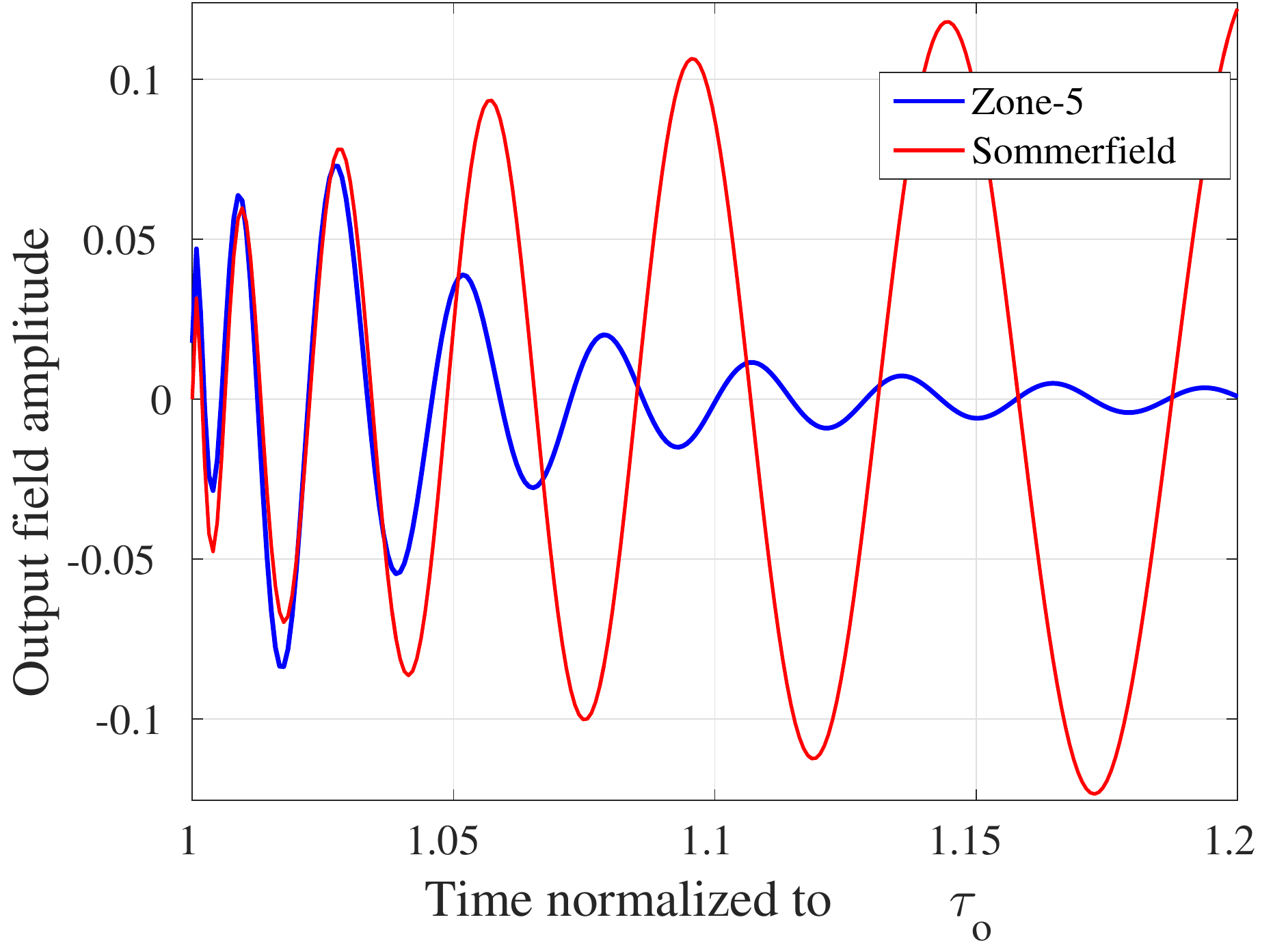}
 \caption[Optional caption]{A comparison between Sommerfeld precursor expression 
 (in red) with the High frequency poles [zone 5] response (in blue).}
 \label{fig:SommerComp}
 \end{figure}

We can approximate the medium permittivity and the slab reflectivity for zone 5 as 
$\sqrt{\epsilon_{\omega}} \simeq 1$ and $\rho_{\omega} \simeq0$.  Assuming 
$|\om_q| \gg |\om_s|$,
the transient solution for the poles of zone 5 given by Eq. (\ref{eq:MainTransient}) 
becomes
\begin{equation} \label{eq:FarTransient}
E_{\text{trans}}^{(5)}(t) =  \frac{-\om_s}{4\pi} \;\sum_{q} 
\frac{ e^{-i \om_q \big[ t-\sqrt{\epsilon(\omega_q)} \big]}}{\om_q^2} \, .
\end{equation}
It is shown in the appendix that this expression can be reduced to the formula 
of the Sommerfeld precursor for a wave 
propagating in a dispersive medium \cite{Bri1960}: 
\begin{equation} \label{eq:Sommer} 
E_{\text{trans}}^{(5)}(t) = \dfrac{\om_s}{4 \pi} \, \dfrac{\sqrt{t - \tau_0}}{\Omega \sqrt{2 \tau_0}}
\: J_1\big( \Omega \sqrt{2 \tau_0} \sqrt{t - \tau_0} \, \big) \, ,
\end{equation}
where $J_1$ is the Bessel function of the first kind.  This formula is valid 
when $(t-\tau_0) / \tau_0 \ll \Omega^2 / 2$. It predicts an initial periodicity of the output 
independent on the main signal frequency $\om_s$. See Appendix \ref{AppB} for more details of 
the derivation of this expression. When compared to the expression (\ref{eq:Sommer}), our 
method presents the advantage that it is not limited to the condition of short times $t-\tau_0$. 
It gives the complete temporal response of the high frequency components of the propagating pulse.
For instance, figure \ref{fig:SommerComp} shows the Sommerfeld precursor for the given test case. 
It raises in the beginning and then gradually decays. The figure compares the derived expression 
for the poles of zone 5 with the Sommerfeld precursor formula which is only valid for the short 
times.
\subsection{Brillouin precursor}
\begin{figure} 
  \includegraphics[width=0.9\linewidth]{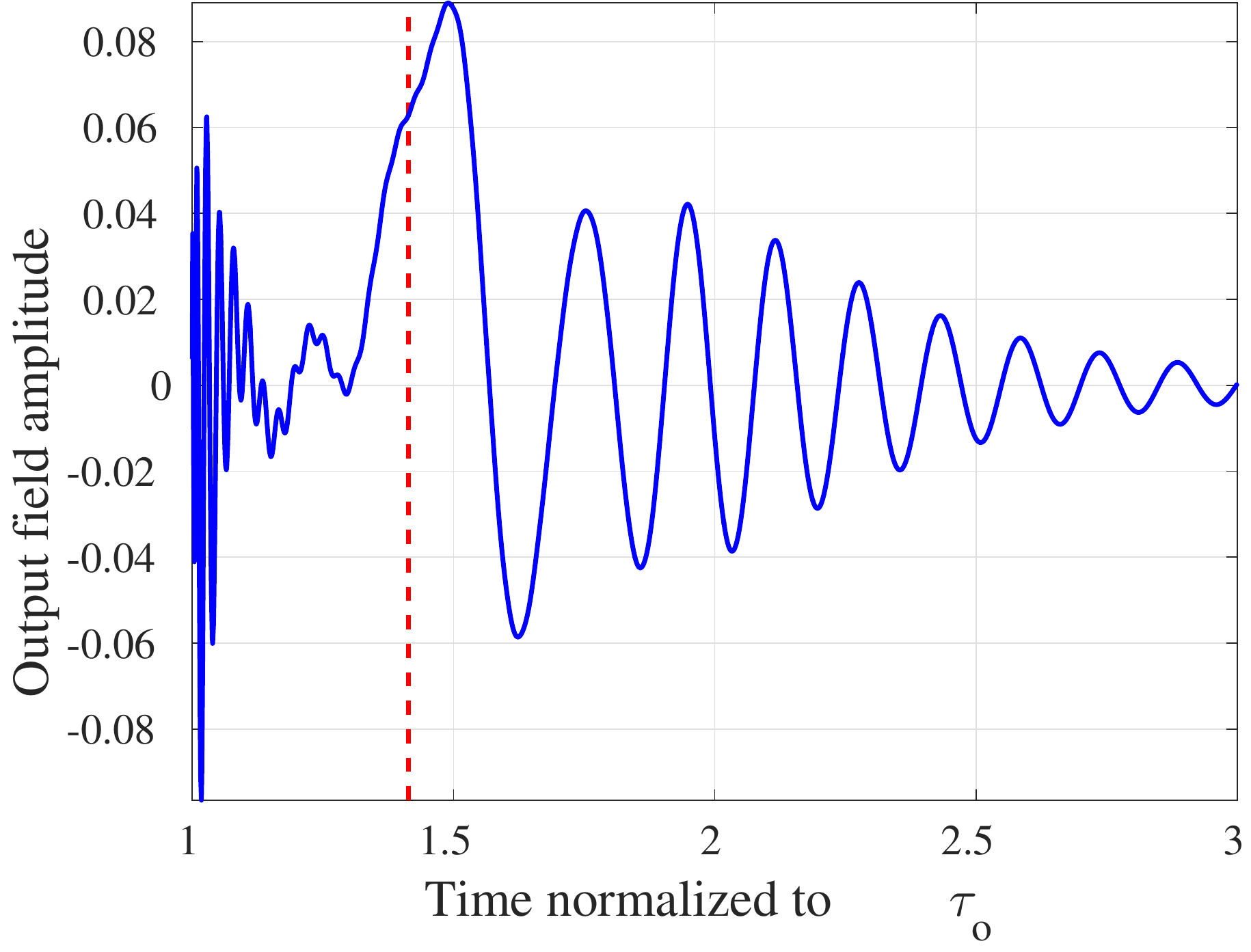}
 \caption[Optional caption]{The precursors of the signal; the rapidly oscillating 
 Sommerfeld followed by the Brillouin precursor that begins from the dotted line.  }
 \label{fig:Precursors}
 \end{figure}

The poles of zone 1 resemble the Brillouin precursor that follows the Sommerfeld one. 
We can use similar assumptions of zone 5 except that the permitivity is given by Eq. 
(\ref{eq:zone1eps}). Figure \ref{fig:Precursors} represents the combined results of zone 1 
and 5 that clearly shows how the rapidly oscillating Sommerfeld precursor followed by they 
slowly oscillating Brillouin precursor. The beginning of the Brillouin precursor is determined 
by the value of $n_g(\om\approx0)= \sqrt{2}$, according to equation (\ref{eq:transtime}).

\subsection{Contribution of the resonance zone}

\begin{figure} 
  \includegraphics[width=0.9\linewidth]{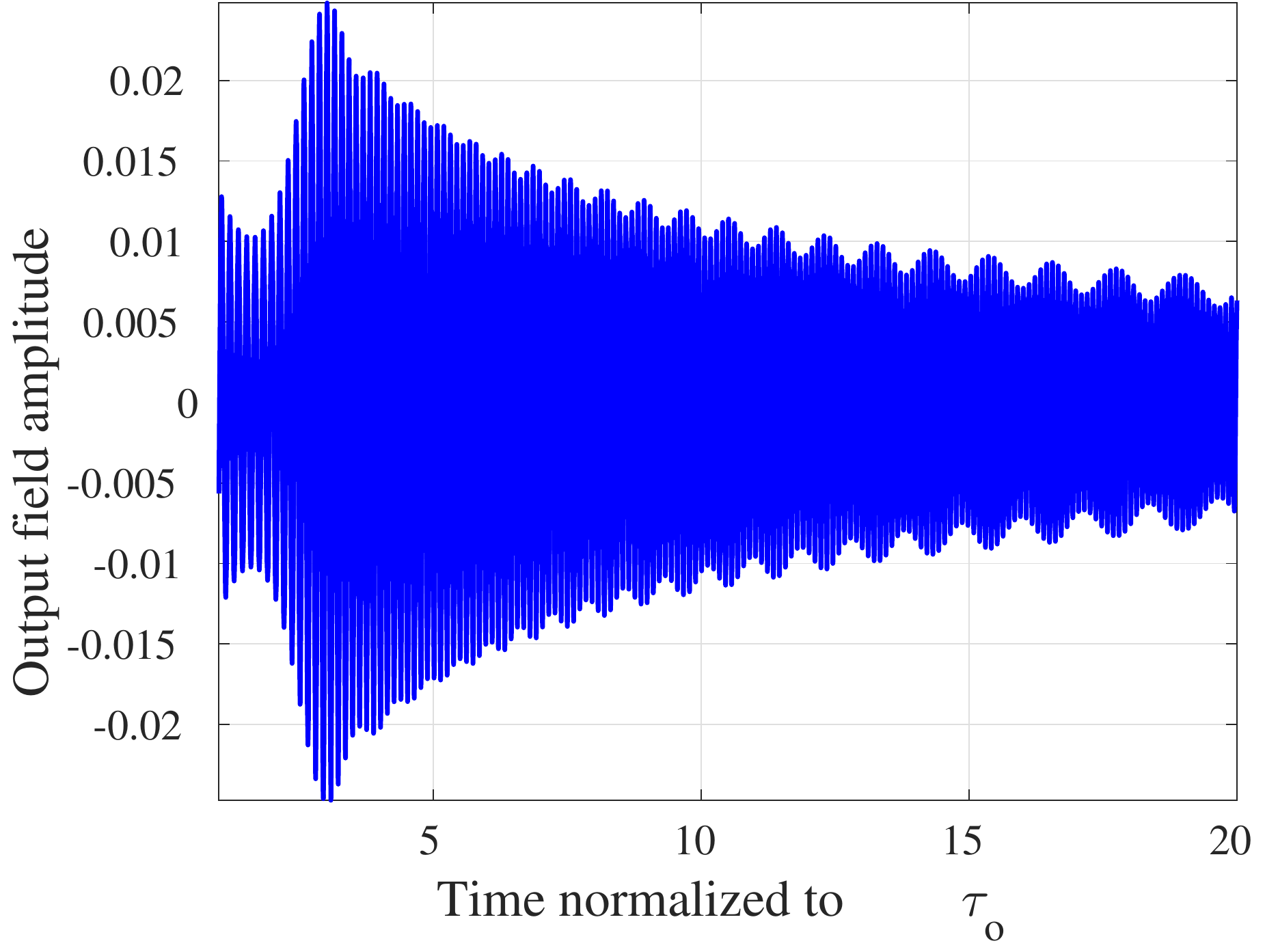}
 \caption[Optional caption]{The temporal response of the near resonance poles.}
 \label{fig:ResPoles}
 \end{figure}

Figure \ref{fig:ResPoles} shows the response of the near resonance poles of zone 2 
for the given test case. Since the excitation frequency is away from the resonance 
region and because the resonance poles have a small imaginary parts, the amplitude 
does not appear significant. 

The effective group velocity of these poles is very small. Therefore, the contribution 
of the resonance poles appears later than other poles. On the other hand, this contribution 
lasts for a longer time because the small imaginary part means that the outcoupling is small, 
and hence the wave at this zone 
can survive for a longer time inside the resonator.

\subsection{The total response}
\begin{figure} 
  \includegraphics[width=0.9\linewidth]{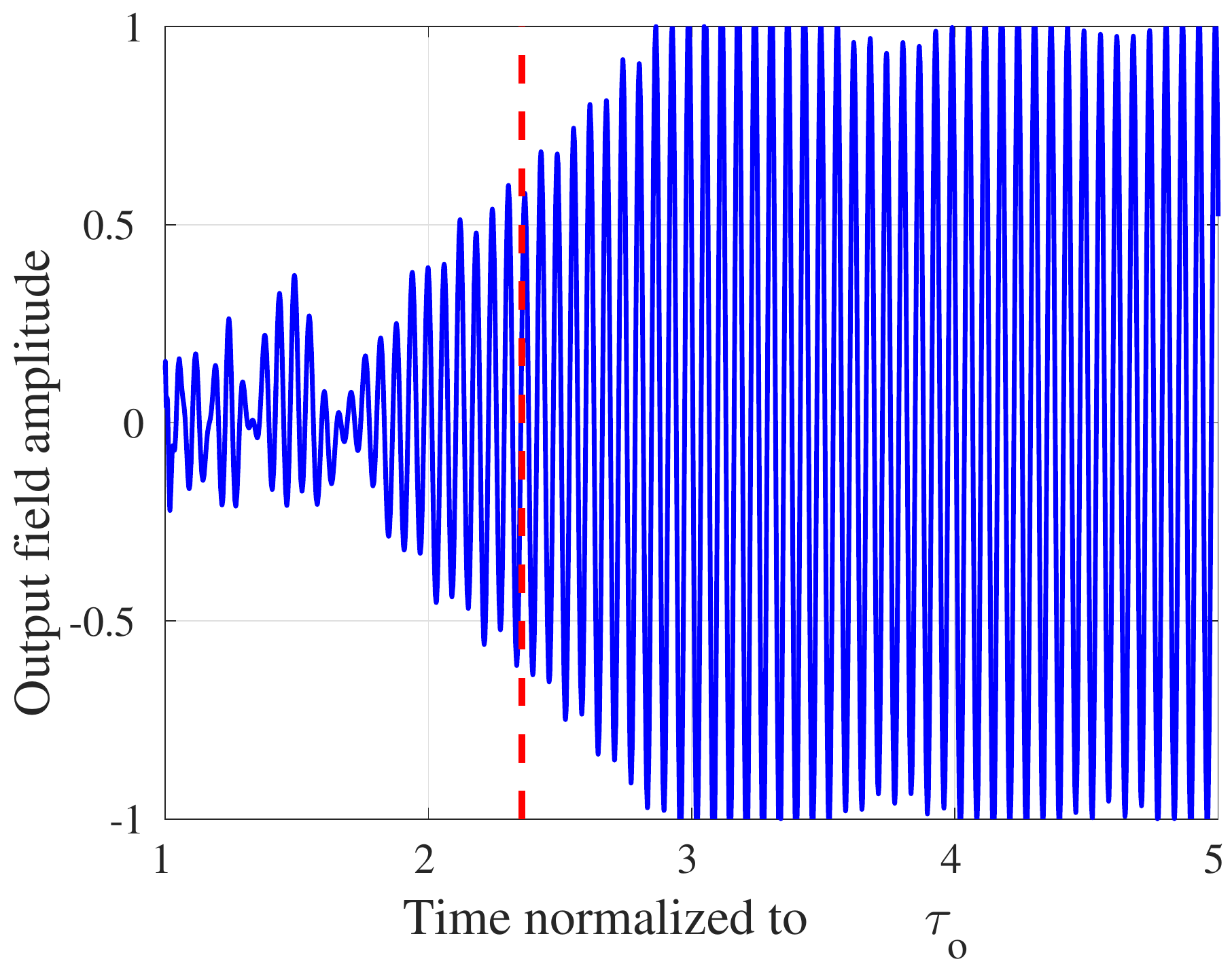}
 \caption[Optional caption]{The total temporal response of the dispersive slab. The dotted line 
 indicates the commencement of the main pulse at the output.}
 \label{fig:totaltemporalresponse}
 \end{figure}

\begin{figure} 
  \includegraphics[width=0.9\linewidth]{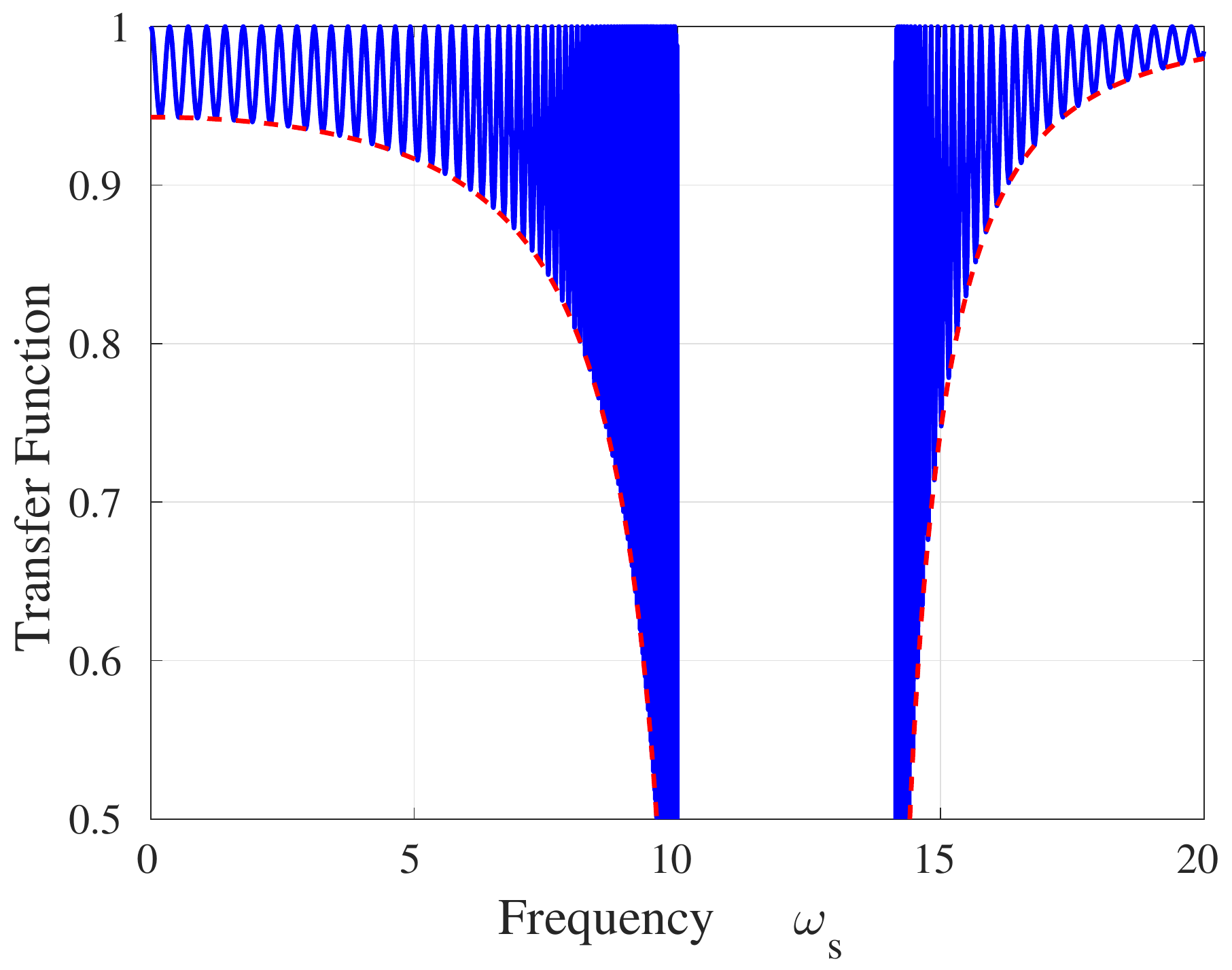}
 \caption[Optional caption]{The transfer function of the dispersive slab (Steady state solution).}
 \label{fig:Stst}
 \end{figure}

The transmitted field from the dispersive slab is given by Eqs. (\ref{twoparts}-\ref{eq:MainTransient}). 
Figure \ref{fig:totaltemporalresponse} shows the total temporal response which includes both 
the steady state and the transient regimes. The steady state value defined as ($t \to \infty$) 
is given by the slab transfer function in Eq. (\ref{eqSlabTF}). 
Figure \ref{fig:Stst} displays the transfer function for the test case. The maximum 
transmission with unity value indicates the position of the poles which are not equally distant 
contrary to the non dispersive case. As we get closer to the plasma 
and resonance zones, the poles get denser. 

The frequency-dependent minimum transmission is also highlighted in the figure by the red lines. 
Its variation is due to the frequency-dependent single interface reflectivity $\rho(\om)$. 
Similarly to Eq. (\ref{eq:Tmin}), this minimum transmission is given by
\begin{equation} \label{eq:Tmindis}
T_{\text{min}} (\omega_s) =  \frac{1-\rho(\omega_s)^2}{1+\rho(\omega_s)^2}
\end{equation}

Since the excitation frequency of the test case is in the plasma zone, the temporal 
response of the poles of zone 4 combined with the source poles forms the main pulse 
at the output. This part of the field starts from a time depending on the group velocity 
at this excitation frequency: its value $n_g(\om_s)=2.36$ is given by Eq. (\ref{eq:transtime}).


\section{Effect of absorption\label{absorp}}
The effect of absorption on the response of a dispersive slab is briefly discussed. The absorption 
means an energy transfer between the incoming wave and the medium. Assuming a small absorption of the 
slab medium ($\gamma \ll \omr$), it can be expected to have a significant absorption only in the 
resonance region. 

The absorption leads to a change in the imaginary part of the poles of the slab. 
The modified  expression for the poles can be obtained near 
the resonance when $\om \approx \omr$ using the lossy expression of the 
permittivity in Eq. (\ref{epDL}) by replacing $\omr$ by $\omr-\i\omr \gamma/2$: in zone 2, 
the poles are then given by 
\begin{equation} \label{eq:PolesLossy}
\omega_{q} \approx \omr -  \frac{ \omr  \Omega^2 }  { 2 q^2 \pi^2 } \left[ 1+i \frac{4\omr}{q^2\pi^2}+ 
i \gamma \frac{q^2}{4\Omega^2} \right] .
\end{equation}

\begin{figure} 	
    \includegraphics[width=0.9\linewidth]{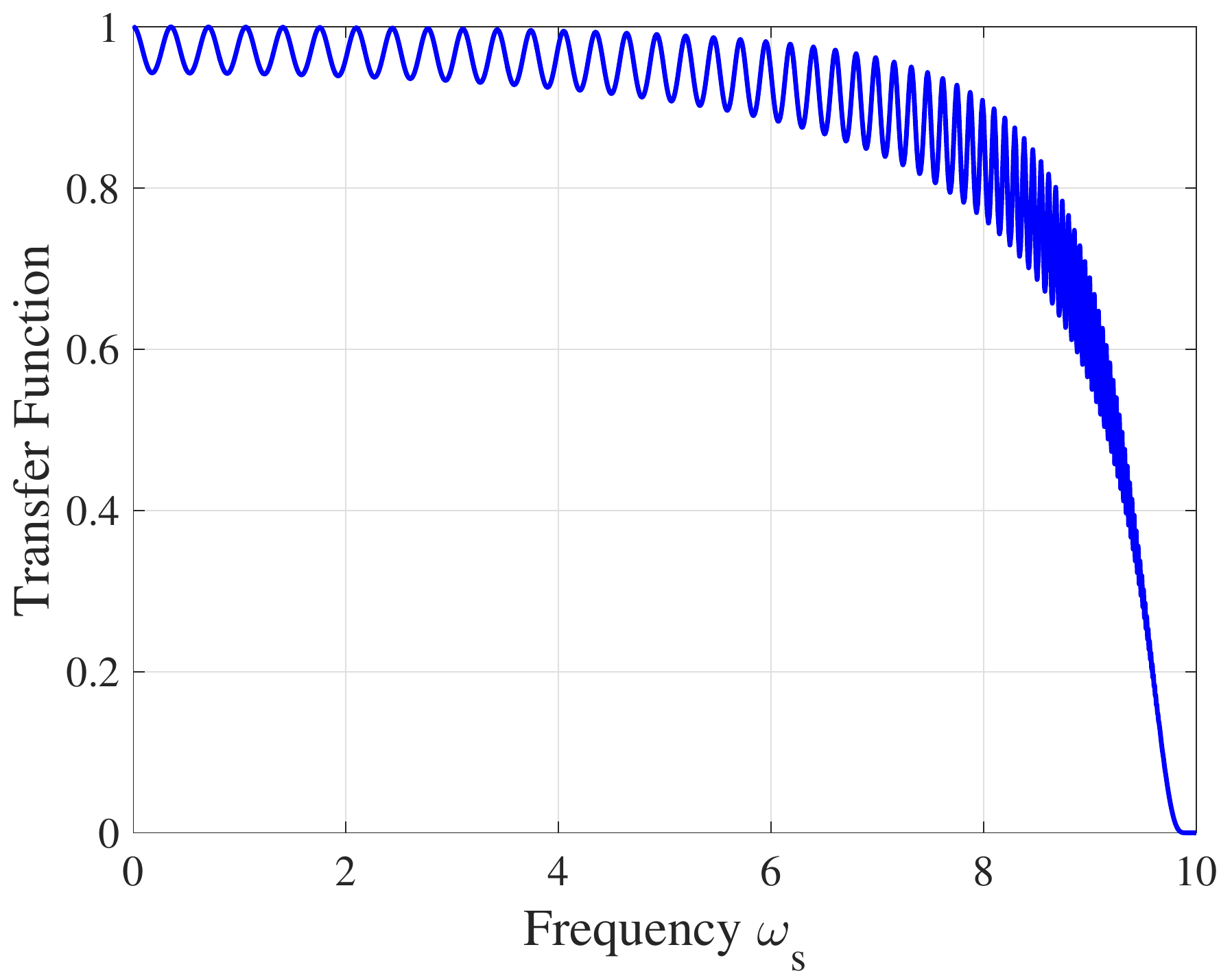}
 \caption[Optional caption]{The transfer function of a lossy slab with $\gamma=0.01$. }
 \label{fig:TFloss}
 \end{figure}

In this case, the imaginary part of the poles is affected by both the medium 
absorption and the reflectivity of the interfaces of the slab. Therefore, for the case of a 
lossy medium, even if the near-resonance excitation frequency matches one of the poles 
of the system, the transfer function is less than unity, see Fig. \ref{fig:TFloss}. 
Depending on the value of the absorption coefficient, the damping of the transient domain 
(imaginary part of the pole) is either governed by the slab outcoupling or by the medium 
absorption. It is stressed that Figure \ref{fig:TFloss} confirms our assumption stating 
that the effect of absorption  mainly occurs at the vicinity of the resonance frequency $\omr$.

\section{Conclusion\label{conclu}}
A general equation has been established for the response of a slab of dispersive medium 
illuminated by a causal sinusoidal electromagnetic plane wave in normal incidence. 
The slab is modeled as a resonator, taking into account the internal reflections at 
the interfaces of the slab with the surrounding medium due to the discrepancy of 
their refractive indices. The advantage of using the resonator model is the 
elimination of the brunch-cut in the complex frequency domain that is usually needed 
to investigate the propagation in a dispersive medium.

The Drude-Lorentz model is then used for the dispersive medium. The poles of the dispersive slab 
have been expressed analytically and then the residue theorem has been used to evaluate the 
temporal response, including both the steady state and transient regimes. A causal pulse 
propagating inside a dispersive medium is shown to split into different temporal regimes 
as each frequency zone of the pulse has its own group velocity. The temporal solutions of
the different frequency zones have been highlighted in a chronological order. The original 
results of Sommerfeld and Brillouin have been retrieved, with the Sommerfeld and 
Brillouin precursors. 

The method proposed in this article appears promising to make progress in the understanding 
of the waves propagation in dispersive media. In particular, the present analysis could be extended 
to the non-normal incidence case, to the absorptive case, and to a dispersion more general 
than the Drude-Lorentz model. 




%



\appendix  
\section{Calculation of poles} \label{AppA}

This Appendix reports the detailed proof of the expressions of the poles of the
 Lorentz-dispersive slab for the frequency different zones, represented in section 
 \ref{sec:poles}. The overall temporal response is consisted of the overall contributions 
 of all types of poles of the slab in addition to the poles of the causal source. 
 Zone 3 has no poles as there is no propagation when $\ep(\om) < 0$. 

It is important to note that the complex frequency domain is symmetric around the 
imaginary axis,  i.e. $\text{Re}(\om_{-q})=-\text{Re}(\om_{q})$ and 
$\text{Im}(\om_{-q})=\text{Im}(\om_{q})$, 
where q is the index labelling the poles. In order to obtain the poles, we need to solve 
the equation of the denominator of the transfer function of the dispersive slab, 
see Eq. (\ref{eq:poleseq}):
\begin{equation} \label{eq:MainApp}
   \rho(\om_q) = \pm 1 \; e^{-i \sqrt{\ep(\om_q)}\; \omega_q}.
\end{equation}
The solution is determined in the form $\om_q= a(q)-ib(q)$, 
where a and b are positive real functions. With this notation, the equation above 
becomes
\begin{equation} 
   \rho(\om_q) = e^{-iq\pi} \; e^{-i a(q) \sqrt{\ep(\om_q)}} \; e^{- b(q) \sqrt{\ep(\om_q)}},
\end{equation}
and, equating the amplitude and the phase parts, it leads to
\begin{equation}
a(q)= \frac{q \pi}{\sqrt{\ep(\om_q)}}.
\label{A:a}
\end{equation}
\begin{equation}
b(q)= - \frac{\ln \big| \rho(\om_q) \big|}{\sqrt{\ep(\om_q)}}.
\label{A:b}
\end{equation}

At each zone a reasonable approximation will be used, which leads to a closed-form expression. 
Since the amplitude term (reflectivity) is bounded to unity, it is generally correct to assume 
that $b \ll a$ and $|\om_q| \simeq a$, except at the static value $q=0$. 
The real part represents the free spectral range of a passive dispersive resonator: 
$\text{FRS} = a(q+1) - a(q)$. 
\subsubsection{[Zone-1]: Low-frequency } \label{sec:DCzone}
Starting from $\om \ll \omr$, Taylor series can be used to approximate the permittivity, and 
the reflectivity at this zone as
\begin{equation}
\ep(\om) \approx \eps + (\eps - 1 ) \, \dfrac{\om^2}{\omr^2} \, , 
\end{equation}
\begin{equation}
\begin{array}{ll}
\rho(\om) & \approx \rho_s \left[1 + \dfrac{1}{2} \, \dfrac{\om^2}{\omr^2} \right]\, . 
\end{array}
\end{equation} 
Since the permittivity is nearly constant in this zone, we assume the FSR  is constant. For this zone,
\begin{equation}
a = \frac{q\pi }{\sqrt{\ep_s}}.
\end{equation}
Using $|\om_q| \simeq a$,
\begin{equation}
b = \dfrac{\ln \big| 1/\rho_{\om_s} \big|}{\sqrt{\eps}} - \dfrac{q^2 \pi^2}{8 \omr^2 \eps \sqrt{\eps}} \, ,
\end{equation}
and the poles expression in zone 1 are,
\begin{equation}
\omega_q = \frac{q\pi }{\sqrt{\ep_s}} - i \dfrac{\ln \big| 1/\rho_{\om_s} \big|}{\sqrt{\eps}}
+ \dfrac{q^2 \pi^2}{8 \omr^2 \eps \sqrt{\eps}}
\end{equation}
This expression is valid as long as $q \ll 2 \sqrt{\ep_s} \omr$.

\subsubsection{ [Zone-2]: Near-Resonance}
For  $\omega \simeq \omr$, we can assume  ($\omega^2-\omr^2) = 2\omr (\omega-\omr)$. 
As a first iteration, let $\omega=\omr - \zeta^2$. In this zone the permittivity has a very large value, 
\begin{equation}\label{epsapprox}
\epsilon(\omega)= 1-\frac{\Omega^2}{\;\omega^2-\omr^2} \simeq \frac{\Omega^2}{2 \omr (\omr-\omega)} \gg 1 \, ,
\end{equation} 
which implies 
\begin{equation} \label{eq:Ressqrteps} 
\sqrt{\ep(\omega)} \simeq   \frac{ \Omega}{\sqrt{2\omr} \sqrt{(\om1-\omega)}}  \simeq   
\frac{ \Omega}{\sqrt{2\omr}} \frac{1}{\zeta} \, ,
\end{equation}
and, assuming $\zeta^2 \ll \frac{\Omega^2}{2\omr}$, we get
\begin{equation}
\rho(\omega) \simeq -1 + \frac{2}{\sqrt{\epsilon(\om)}} = - 1 + 2 \frac{\sqrt{2\omr}}{\Omega} \, \zeta \, .
\end{equation}
Taking the logarithm of the square of Eq. \ref{eq:MainApp} yields
\begin{equation}
- 4 \frac{\sqrt{2\omr}}{\Omega} \zeta = i q 2 \pi - i 2 \dfrac{ \Omega}{\sqrt{2\omr}} \, \dfrac{1}{\zeta}\, (\omr-\zeta^2).
\end{equation}
Since $\zeta^2  \ll \frac{\Omega^2}{2\omr}$, the left side can be ignored. 
In order to reach an expression for $\zeta$ for the first iteration we assume in addition 
$ \zeta^2 \ll \omr$, which leads to
\begin{equation}\
\zeta=   \frac{ \sqrt{\omr / 2}  \Omega }  {  q \pi }.
\end{equation}
Consequently, the poles expression is, after a first iteration, 
\begin{equation}\
\omega_{q} = \omr -  \frac{ \omr  \Omega^2 }  { 2 \pi^2 q^2  }, 
\end{equation}
where $q \gg  \Omega , \om_o$.
As expected, no pole exists higher than the resonance where the refractive 
index is complex (metallic region).

To obtain the imaginary part of the poles expression, a second iteration is needed: 
\begin{equation} \label{eq:respoleassm}
\omega_{q} = \omr -  \frac{ \omr  \Omega^2 }  { 2 \pi^2 q^2  } (1+\zeta^2).
\end{equation}
Here, it is assumed that $\zeta^2 \ll 1$. Similarly to the first 
iteration, we have
\begin{equation}\
\ep(\om_q)\simeq   \frac{\pi^2 q^2}{ \omr^2 (1+\zeta^2)},
\end{equation}
and
\begin{equation}\  \label{eq:rhoResN}
\rho(\om_q)  = - 1 +  \frac{2\omr}{q \pi} \left[ 1+\frac{\zeta^2}{2} \right] \, .
\end{equation}
Taking again the logarithm of the square of Eq. \ref{eq:MainApp}, 
it is obtained that
\begin{equation}
\begin{array}{ll}
- \dfrac{4\omr}{q \pi} \left[ 1+\dfrac{\zeta^2}{2} \right] = \\[4mm]
\qquad \qquad i q 2\pi - i \dfrac{ 2\pi q}{\omr} 
 \left[ 1-\dfrac{\zeta^2}{2} \right]  \left[ \omr -  
 \dfrac{ \omr  \Omega^2 }  { 2 \pi^2 q^2  } (1+\zeta^2) \right] \, .
 \end{array}
\end{equation}
Using that $q \gg  \Omega$ and $\zeta^2 \ll 1$, we reach
\begin{equation}\label{eq:zetares}
- \dfrac{4\omr}{q \pi} \approx i q \pi \zeta^2
\quad \Longrightarrow \quad \zeta^2 = i \frac{4 \omr}{q^2\pi^2}.
\end{equation}
The expression of poles obtained after two iterations is then
\begin{equation}\ \label{nearres}
\omega_{q} = \omr -  \frac{\omr  \Omega^2 }  {2  q^2 \pi^2 } \left[ 1 +i \frac{4\omr}{q^2\pi^2} \right].
\end{equation}
This expression is limited to $q \gg  (\omega_o,\Omega).$
\subsubsection{[Zone 4]: Near Plasma frequency } \label{sec:AppPlas}
For near-plasma poles, the parameter $\zeta^2 \ll 1 $ is introduced:
\begin{equation}\label{epsapp}
\omega_{q} = \omega_p ( 1+ \zeta^2). 
\end{equation} 
where $\omega_p = \sqrt{\omr^2 + \Omega^2}$ is the plasma frequency. Then the permittivity becomes
\begin{equation}\label{epsapp}
\ep(\omega)  \simeq \frac{2 \omega_p^2}{\Omega^2} \, \zeta^2  \quad 
\Longrightarrow \sqrt{\ep(\omega)} \approx \frac{\sqrt{2} \omega_p}{\Omega} \, \zeta 
\end{equation} 
and, using that this permittivity is near zero, we get
\begin{equation}\label{epsapp}
\rho(\om) \approx 1 - \frac{2\sqrt{2} \omega_p}{\Omega} \, \zeta  \, .
\end{equation} 
Taking the logarithm of  Eq. \ref{eq:MainApp} leads to
\begin{equation}\label{epsapp}
- \frac{2\sqrt{2} \omega_p}{\Omega} \, \zeta = i q \pi - i \frac{\sqrt{2} \omega_p^2}{\Omega} \, \zeta ( 1+ \zeta^2) \, ,
\end{equation}
and, using the presumption of $\zeta^2 \ll 1$, it is obtained that
 \begin{equation}\label{epsapp}
\zeta = \frac{q \pi \Omega}{2\sqrt{2} \omega_p (1 + \omega_p^2/2)} (\omega_p / 2 - i) \, .
\end{equation}
To simplify the expression and since the plasma wavelength of materials are mostly in the 
nanometer region \cite{DresselEM}, then the slab thickness can be assumed to be much larger 
than the plasma wavelength. In this situation, the normalized plasma frequency is $\omega_p \gg 1 $. 
Finally, we reach an expression for the poles near the plasma frequency $\omega_p$,
 \begin{equation}\label{epsapp}
\omega_{q}=\omega_p \; \left[ 1+ \frac{q^2 \pi^2 \Omega^2}{8 \omega_p^4}  \;(1 - 4 i / \omega_p) \right].
\end{equation}  
The limit of validity for this expression is $q \ll \frac{\omega_p^2}{ \Omega}$ 
in order to satisfy $\zeta^2 \ll 1$.

\subsubsection{[Zone 5] High Frequency } \label{sec:AppFar}
In this zone it can be used that $\epsilon(\omega)\simeq 1$, with
\begin{equation}\label{sqrteps}
\sqrt{\epsilon(\omega)} \simeq 1 - \frac{\Omega^2 /2 }{\omega^2-\omr^2} \, ,
\end{equation}
which implies
\begin{equation}\label{rho}
\rho(\omega)  \simeq \frac{\Omega^2 / 2}{\omega^2-\omr^2}  \simeq 0,
\end{equation}
Hence, from (\ref{A:a}) and (\ref{A:b}), it is obtained that
\begin{equation}\label{ab-parts}
a(q) = {q\pi} \, , \quad
b (q)=  \ln{\frac{q^2 \pi^2 -\omr^2}{\Omega^2/2}}.
\end{equation}
The poles expression is then
\begin{equation}\label{eq:zone5poles}
\omega_{q} = {q\pi} - i \ln{\frac{q^2 \pi^2 -\omr^2}{\Omega^2/2}} \, .
\end{equation} 
This expression for frequency poles remains valid as soon as $\pi^2 q^2 \gg \omr^2 + \Omega^2 /2 $. 
\section{Derivation of Sommerfeld precursor} \label{AppB}


We start with eq. (\ref{eq:FarTransient}):
\begin{equation}
E_{\text{trans}}^{(5)}(t)=  - \frac{\om_s}{4\pi} \;\sum_{q} 
\frac{ e^{-i \om_q \big[ t-\sqrt{\epsilon(\omega_q)} \tau_0 \big]}}{\om_q^2} \, .
\end{equation}
Let $\Delta = t - \tau_0$ with $\tau_0=1$ normalized to unity. Assuming 
$\om \gg \omr$, Eq. \ref{sqrteps} can be used,
\begin{equation}
\sqrt{\epsilon(\om)}= \sqrt{1- \frac{\Omega^2}{\om^2-\om_o^2}} \approx 1-\frac{\Om^2}{2 \om^2},
\end{equation}
that leads to
\begin{equation}
E_{\text{trans}}^{(5)}(t) \approx - \frac{\om_s}{4\pi} \;\sum_{q} \frac{ e^{-i \om_q  \Delta -i \frac{\sigma}{\om_q}}}{\om_q^2},
\end{equation}
with $\sigma = \Omega^2 \tau_0 /2$. In order to turn it into an integral form over the real axis 
in the complex domain so that it can be compared with Sommerfeld expression \cite{Bri1960}, the 
following scaling is considered: let $\om_q'=\om_q \alpha,\; \sigma'=\sigma \alpha$ and 
$\Delta'=\Delta / \alpha$. Setting $\alpha$ close to zero allows us to convert the 
summation into an integration over the integrand, $d\om_q' = \pi \alpha dq' $ which is given 
from Eq. (\ref{eq:zone5poles}),
\begin{equation}
E_{\text{trans}}^{(5)}(t) \approx - \frac{\alpha \om_s}{4 \pi^2} \int_{\mathbb{R}} 
\frac{d\om_q'}{{\om_q'}^2}   e^{-i \om_q'  \Delta' 
- i \frac{\sigma'}{\om_q'}}.
\end{equation}
For $t \to \tau_0$, the parameters $\Delta$ and $\Delta'$ can be arbitrary small, and the 
integral above over the real axis can be approached by an integral over a semi circle in 
the upper half plane of complex frequencies $\om_q'$ with radius $\sqrt{\sigma' / \Delta'}$ 
arbitrary large: hence the following variable of integration is considered with $u \in [0,\pi]$
\begin{equation}
e^{iu} = \om_q' \sqrt{\frac{\Delta'}{\sigma'}} \, ,
\end{equation}
and the transient field becomes 
\begin{equation}
E_{\text{trans}}^{(5)}(t) = - \frac{\alpha \om_s}{4 i \pi^2} \sqrt{\frac{\Delta'}{\sigma'}} 
\int_{[0,\pi]}e^{i \big[ u -2 \sqrt{\Delta'\sigma'}cos(u)} \big] \; du \; .
\end{equation}
Notice that the result is independent of the scaling $\alpha$. Finally, using the 
expression of Bessel function of first kind
$J_1$, the following expression id obtained:
\begin{equation}
E_{\text{trans}}^{(5)}(t) \approx \dfrac{\om_s}{4 \pi} \, \dfrac{\sqrt{t - \tau_0}}{\Omega \sqrt{2 \tau_0}}
\: J_1\big( \Omega \sqrt{2 \tau_0} \sqrt{t - \tau_0} \, \big) \, .
\label{Sommerfeld-precursor}
\end{equation} 
\begin{acknowledgments}
We would like to show our gratitude to Aladin Hassan Kamel (Visiting Professor at Ain-Shams 
University, Egypt) for the valuable discussions regarding the definition of causality and 
the application of the residue theorem in the case of propagation in a dispersive medium. \end
{acknowledgments}

%

\end{document}